\newif\iflncs
\newif\ifdgruyter
\newif\ifanonymous

\lncstrue

\newif\ifonecolumn
\newif\iftwocolumn
\iflncs
    \onecolumntrue
\fi
\ifdgruyter
    \twocolumntrue
\fi

\ifdgruyter
    \documentclass[USenglish,oneside,twocolumn]{article}
\fi
\iflncs
    \documentclass[11pt]{llncs}
\fi

\usepackage{preamble}

\ifdgruyter
    \DOI{foobar}
    \cclogo{\includegraphics{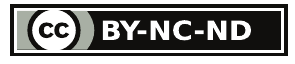}}
\fi

\begin{document}
\ifdgruyter
    \author[1]{Georgia Avarikioti}
    \author[2]{Roman Brunner}
    \author[3]{Aggelos Kiayias}
    \author[4]{Roger Wattenhofer}
    \author[5]{Dionysis Zindros}

    \affil[1]{ETH, zetavar@ethz.ch}
    \affil[2]{ETH, robrunne@student.ethz.ch}
    \affil[3]{University of Edinburgh, IOHK, aggelos.kiayias@ed.ac.uk}
    \affil[4]{ETH, wattenhofer@ethz.ch}
    \affil[5]{University of Athens, IOHK, dionyziz@di.uoa.gr}

    \title{\huge Structure and Content of the Visible Darknet}

    \runningtitle{Structure and Content of the Visible Darknet}


    \journalname{Proceedings on Privacy Enhancing Technologies}
    \DOI{Editor to enter DOI}
    \startpage{1}
    \received{..}
    \revised{..}
    \accepted{..}

    \journalyear{..}
    \journalvolume{..}
    \journalissue{..}
\fi
\iflncs
    \author{Georgia Avarikioti\inst{1} \and
            Roman Brunner\inst{1} \and
            Aggelos Kiayias\inst{2} \and \\
            Roger Wattenhofer\inst{1} \and
            Dionysis Zindros\inst{3}}

    \institute{ETH \and
               University of Edinburgh, IOHK \and
               University of Athens, IOHK}

    \title{Structure and Content of the Visible Darknet}
    \maketitle
\fi

\begin{abstract}
{
In this paper, we analyze the topology and the content found on the ``darknet'', the set of websites accessible via Tor. We created a darknet spider and crawled the darknet starting from a bootstrap list by recursively following links. We explored the whole connected component of more than 34,000 hidden services, of which we found 10,000 to be online. Contrary to folklore belief, the visible part of the darknet is surprisingly well-connected through hub websites such as wikis and forums. We performed a comprehensive categorization of the content using supervised machine learning. We observe that about half of the visible dark web content is related to apparently licit activities based on our classifier. A significant amount of content pertains to software repositories, blogs, and activism-related websites. Among unlawful hidden services, most pertain to fraudulent websites, services selling counterfeit goods, and drug markets.
}
\end{abstract}

\ifdgruyter
    \maketitle
\fi
\iflncs
  \thispagestyle{plain}
\fi
\section{Introduction}

 
Crawlers and search engines have fundamentally changed the web. In June 1993, the first crawler-based search engine Wandex knew just 130 web sites. Soon after, the world witnessed the emergence of many more search engines such as 
Excite, Lycos, AltaVista, and Google. The usability and success of these search engines may have fueled the exponential growth of the web in the mid-1990s.


Today, many web pages offer content that is dynamically generated for each visit, and as such difficult to crawl. Also, a large part of the web is hidden from crawlers: web sites that dislike to be crawled can include a \texttt{robots.txt} file, and a large part of the web is protected behind passwords. All these web sites (dynamic content, password- or robot-protected) are known as the \emph{invisible} part of the web.


While interactions between users and web sites became increasingly encrypted (https), the very nature of the internet protocol (IP) cannot prevent that one can observe who is communicating with whom, e.g. whether a user visits a certain web page.


Like the regular web (the ``clearnet''), the Tor darknet also has a large invisible part (dynamic content or password protected). Because names of web servers are generally non-mnemonic, being invisible in the darknet is the default rather than the exception.


On the other hand, the darknet also features web servers that absolutely want to be visible. The most prominent example is facebookcorewwwi.onion, an anonymous entrance to the regular Facebook web site. More surprisingly, even if you are in an unlawful business (e.g. selling drugs, guns, or passports), you may want to be visible, as there is no business for those who are completely invisible. Figure~\ref{tab:overview} summarizes all these possible web site types.


\ifdgruyter
  \begin{table}[h!]
    \begin{center}
      \caption{Overview of different web site types. The deep web consists of the italicized cells.}
      \label{tab:overview}
      \begin{tabular}{c|c | c }
        & \textbf{Clearnet} & \textbf{Darknet}\\
        \hline
        \hline
        \textbf{Visible}&
        \shortstack[c]{Googleable web \\ Publicly accessible sites} & 
        \shortstack[c]{\\[1pt] \textit{[our paper]} \\ \textit{Visible part of the Darknet}}\\
        \hline
        \textbf{Invisible}&
        \shortstack[c]{\\[1pt] \textit{Your e-mail inbox} \\ \textit{Protected networks}} & 
        \shortstack[c]{\textit{Where spies meet?} \\ \textit{Dark side of the Darknet}}\\
      \end{tabular}
    \end{center}
  \end{table}
\fi

\iflncs
  \begin{table}[h!]
    \begin{center}
      \begin{tabular}{c|c | c }
        \hline
        & \textbf{Clearnet} & \textbf{Darknet}\\
        \hline
        \hline
        \textbf{Visible}&
        \shortstack[c]{Googleable web \\ Publicly accessible sites} & 
        \shortstack[c]{\\[1pt] \textit{[our paper]} \\ \textit{Visible part of the Darknet}}\\
        \hline
        \textbf{Invisible}&
        \shortstack[c]{\\[1pt] \textit{Your e-mail inbox} \\ \textit{Protected networks}} & 
        \shortstack[c]{\textit{Where spies meet?} \\ \textit{Dark side of the Darknet}}\\
        \hline
      \end{tabular}
      \vspace{10pt}
      \caption{Overview of different web site types. The deep web consists of the italicized cells.}
      \label{tab:overview}
    \end{center}
  \end{table}
\fi

%
So does it make sense to build a crawler and search engine for the visible part of the darknet, a ``Google for the darknet?'' While it is generally believed that the darknet is impenetrable~\cite{Biryukov2014,Nabki2017}, we show that the darknet has a surprisingly substantial visible part (Section \ref{sec:data}); moreover, we show that this visible part of the darknet is well-connected with hyperlinks, and that crawling indeed is possible. In addition, in Section \ref{sec:results}, we use machine learning techniques to understand the content of the visible darknet: language, topic of content (e.g. drugs, blogs, adult etc), and legality of content.  
We compare our results to previous work in Section \ref{sec:relatedwork}.

Our contributions are summarized as follows:

\begin{enumerate}
    \item We develop and publish novel open source implementations of a darknet crawler and classifier which addresses unique challenges not typically encountered in clearnet settings.
    \item We provide the first topological analysis of the Tor network graph, concluding that it is well connected and scale-free.
    \item We apply supervised machine learning on downloaded content and report our findings on physical language, topic and legality.
\end{enumerate}

\section{Crawling the Darknet}

\ifdgruyter
\begin{figure}[H]
\includegraphics[width=\linewidth]{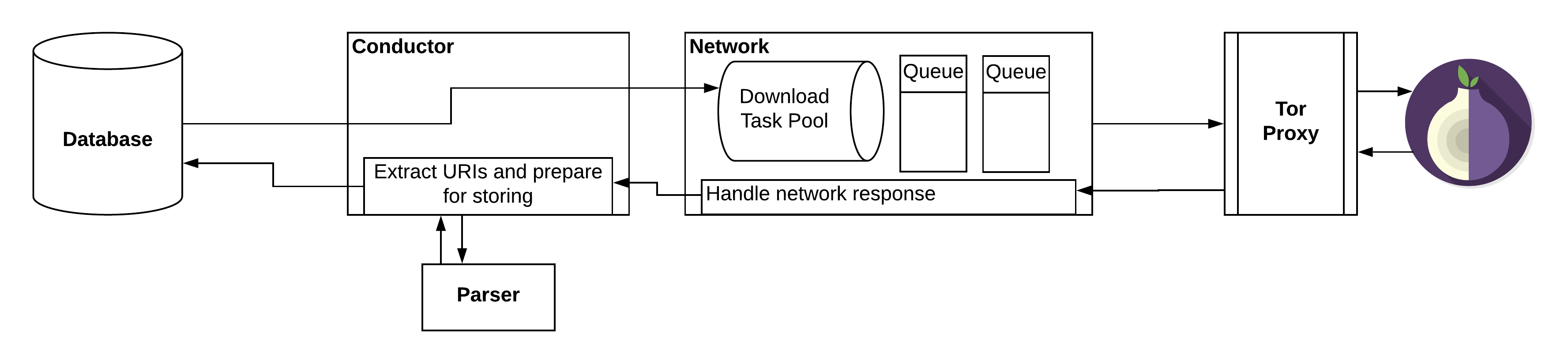}
\caption{The architecture of the crawler. Modules are depicted in bold.~\cite{Loconte}}
\label{fig:scraperArchitecture}
\end{figure}
\fi

\iflncs
\begin{figure}[H]
\centering
\includegraphics[width=\linewidth]{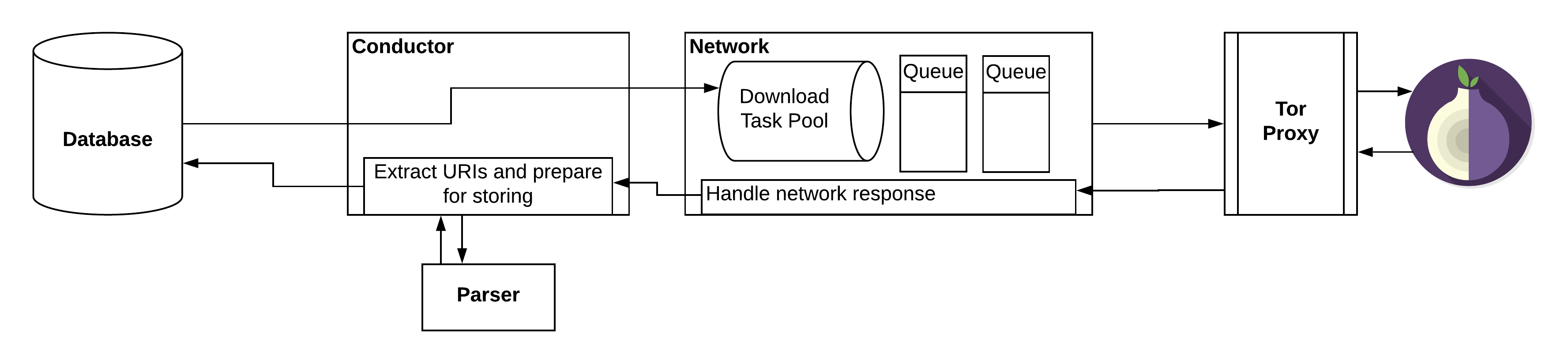}
\caption{The architecture of the crawler. Modules are depicted in bold.~\cite{Loconte}}
\label{fig:scraperArchitecture}
\end{figure}
\fi
In this section, we describe the structure of the spider we built to crawl the visible darknet. To understand some of our design choices, we present the encountered problems and the different strategies applied to resolve these issues. 

The goal of the crawler is to discover and explore a large part\footnote{We measure the size of the darknet in number of hidden services.} of the darknet by recursively following any links it finds in the downloaded content. We distinguish a \emph{hidden service}, which refers to a particular \texttt{.onion} address, from a \emph{path}, which is a URL of a particular website hosted on a specific hidden service.  We store the downloaded content itself, the extracted URIs, the link structure, as well as some metadata such as timestamps and data type of the content.
We seeded our crawler with a list of 20,000 hidden service addresses from previous work \cite{Kadianakis2017}, which allowed us to reach a high initial crawl speed. Out of those 20,000 hidden services, only approximately 1,500 were still online. 

Our crawler uses the following four modules:
\begin{itemize}
    \item \textbf{Conductor:} Manages and orchestrates the download, processing, and storage of content by scheduling the other components.
    \item \textbf{Network module:} Downloads content and handles network related exceptions and issues.
    \item \textbf{Parser:} Parses the downloaded content and extract URIs that will be used recursively to download further content.
    \item \textbf{Database:} Stores the downloaded content as well as the network's structure and URIs.
    \item \textbf{Tor proxy:} Enables the crawler to access the Tor network. 
\end{itemize}

The modules are depicted in Figure~\ref{fig:scraperArchitecture}; they are described below in detail.

The crawler is newly developed for this work and made public under the MIT license for further use by other researchers\footnote{\url{https://github.com/decrypto-org/spider}}. It implements a complete data analysis chain, from collection to data preprocessing and classification. The data collected not only contains the content of the hidden services but also the link structure, status information, network topology information and can be configured to track the network changes over time.
%
%
\subsection{Downloading the Content}
\label{sec:MIMEType}
The network module handles any network related events during the crawling process. It takes a \emph{download task} as input. Such a download task consists of a base URL and a path as well as a unique ID. After getting a download task, a request is sent to the \emph{Tor proxy} \cite{Boyd}. As soon as the network module gets the first data of the response, it starts checking and filtering the contents for potential illegal media content. If the content passes the filter, found to be allowed, and the HTTP status code does not indicate an error, we wait for the download to finish and add the content to the result object. This result object then is passed back to the conductor for further processing.

\sh{Speed Issues of the Tor Network} 
The Tor network is slower than the clearnet \cite{TorMetricsPerformance} and bursty traffic over one single Tor circuit easily congests at least one of the nodes. To prevent overloading of single nodes and to increase the crawling speed, we use a configurable amount of Tor instances in our Tor proxy~\cite{Boyd}. The incoming requests to the Tor proxy are then distributed across all instances of Tor using a Round-robin scheduler. This prevents congesting single nodes and increases download speed dramatically. We found that using 100 instances, we can download a maximum of 30,000 pages per hour.

The download from the Tor network is the slowest part of the crawler, even when employing a large number of Tor instances. In order to achieve the highest possible download speed, we want to keep the downloads running without having to wait for the other modules. Towards this, we introduced a \emph{pool} of download tasks such that the network module never has to wait for the database to deliver new tasks. The pool is of configurable size and is repopulated by the \emph{conductor}. This pool resides in memory to give the network module fast access to new download tasks. If the pool's size hits a configured lower bound of entries, it is repopulated with new download tasks gathered from the database.

The pool size can be configured to fit the machines' capabilities. A system using an SSD can work with a smaller pool, since it takes less time to repopulate the pool compared to an HDD. This results also in a smaller memory footprint of the crawler.

\label{sec:DoSLike}
\sh{Preventing DoS-like Behaviour}
Early in the process we discovered an issue when just blindly downloading anything that is passed on as a download task. If a large part of the download tasks point to the same hidden service and the network module opens up a connection to the same hidden service for each of those download tasks, a darknet web server might classify our behaviour as an attempted DoS attack. First, we do not want to disturb the targeted services. Secondly, servers may initiate rate limiting countermeasures against what they consider a DoS attack, e.g. captchas or similar barriers.

In order to prevent this, we implemented a \emph{push back mechanism}. This mechanism keeps track of the number of already running requests per hidden service and only starts a new download task to the same hidden service if there are less than four live concurrent connections to this hidden service. If all available slots for this hidden service are in use, the download task is put into a waiting queue. This queue will be checked for waiting requests as soon as one of the four pending requests finishes.

The value of four concurrent requests was chosen to be lower than the default Tor Browser's setting of six\footnote{Six concurrent connections is the default in Tor version 7.5.6. The setting is found under \texttt{network.http.max-persistent-connections-per-server}.}. Being below makes the crawler less detectable. Note that four concurrent requests is higher than the proposed number of two simultaneous connections by RFC 2616 \cite{Nielsen1999}; however, most modern browsers use a value between four and eight as well.

The entries in the queues do need memory. If the pool is repopulated and a lot of download tasks reside in the queue, the memory consumption grows unboundedly. This issue cannot be solved by the network module but requires the other modules that repopulate the pool to choose the entries in the pool wisely, such that not too many download tasks point to the same hidden services and therefore end up in the waiting queue.

Additionally, many concurrent download tasks in the waiting queues can slow down the crawling progress.
To ensure the queues will be emptied eventually we do the following every time a download task returns: we
check whether there is a download task in a waiting queue that does not violate our limit of four connections per hidden service. If such a task exists we pop it out from the queue, else we get a new entry from the pool. However, in a worst case scenario all hidden services in progress have four concurrent tasks from the waiting queues. Then, all available connections will be busy emptying these queues and no further progress will be made towards other hidden services. This implies that the number of hidden services needed to delay the overall progress of the crawler on the Tor network is $25\%$ of the total number of connections.

\sh{Storage and Processing of Illegal Content}
Since we download any URI we find on the darknet, the legality of the content is unknown before inspection. To prevent storing illegal data such as child pornography persistently, we apply filtering at different stages. The goal of these filters is to only store text. Towards this end we employ a series of black- and whitelist filters.

First we filter by the returned MIME type, if available. This allows us to only whitelist supported MIME types, in our case any textual types. This happens at the earliest possible stage of filtering, right after we have received the response header. We need to flush our cache if we detect an unsupported MIME type, since potentially illegal content might already be in memory. The only information we keep in this case is the HTTP status code returned, the MIME type and the timestamp when the answer was recorded.

If the MIME type header is not set, we need to download the full content to decide which MIME type fits by content inspection using mmmagic~\cite{mscdex}. This requires the full data in memory, but this data will also be immediately discarded if the detected MIME type is not whitelisted.

After the whitelist MIME type filter, only text documents should be available. These can still contain textual representations of media content in the form of data URIs such as base64 encoded data. Here we employ a blacklist filter that replaces any occurrence of a non-textual data URI by a placeholder. 

The previous steps all happen in memory and that data is not accessible to anybody. At this point, all rich media content should be removed, and therefore the content is ready to be passed back for persistent storage. 
%
%
\subsection{Orchestrating the Crawling}
The conductor balances the needs of the other modules. It is the conductor's responsibility to dispatch URIs for download to the network module and get new URIs from the database into the download task pool of the network module. It is important for the conductor to keep the number of download tasks in the pool balanced in order to prevent an unpredictable increase in memory usage. It also uses information about the network module queues when repopulating the pool, by only requesting paths from hidden services that have not yet entries in their queue.

\label{sec:prioritizationDownload}
\sh{Prioritization of Downloads}
Deciding which URI to schedule next for download is important not only to prevent DoS-like behaviour as described in Section \ref{sec:DoSLike}, but also to ensure we make progress on the discovery of the visible part of the darknet. We applied several prioritization schemes to ensure good progress and to prevent issues with some pathological examples. We will now introduce those different schemes and explain for each why we introduced them, when they work and where they fail. 

\begin{itemize}
  \item \textbf{Na\"ive}: The first approach was to greedily fill up the pool with new download tasks in the order they were previously inserted into the database. However, most of the pages do link a lot to pages of the same hidden service and contain only a small amount of links that point to different hidden services. When populating the pool with such entries, we will fill up immediately the queues in the network module and increase memory consumption. In that case, the speed of exploration of the darknet might be slowed down, as the network module is busy processing the queues.
  \item \textbf{Random}: To prevent this issue, we started randomizing the selection of paths inserted into the download task pool. This works well when all hidden services in the database have a similar number of paths available for download. However, there exist pages that contain a huge number of links to themselves -- we call these \emph{black hole pages}. An example for a black hole page is a bitcoin explorer that lists every action ever recorded on the bitcoin blockchain. This explorer took up to $90\%$ of the available paths in our database in an early version of our crawler. Even randomization then will lead to a pool filled with $90\%$ download tasks pointing to the same hidden service. The random strategy also does not do a good job at discovering new hidden services, since we select the next page to be downloaded at random.
  \item \textbf{New-hidden-service-first}: This strategy tries to advance through the darknet as fast as possible by downloading only a single page per hidden service. This method is efficient when the collection of yet uncrawled hidden services is large because it advances very quickly to new hidden services. However, the first page downloaded from a hidden service is not always useful for classification or collection of further links. Thus this strategy only works well for seed generation but not for data collection.
  \item \textbf{Prioritized}: Since the \emph{new-hidden-service-first} strategy does not guarantee good data for link collection or classification, we attempted to use another part of the dataset we collected along the crawling progress to serve as an indicator of which path should be dispatched next. For each path, we calculate the number of unique hidden services linking to this path. This count then serves as a proxy to assess the importance of a path~\cite{PageRank}. When many unique hidden services link to a particular path, that path likely contains important information. We then start downloading the path with the highest distinct incoming link count. Although this strategy works well in general and results in a nice ordering of paths to be scheduled, there exist two issues with this strategy. First, we found that the darknet contains closely interconnected subnetworks. One such subnetwork is a bitcoin scamming scheme. The scammers spin up several hundred hidden services, which then link to each other in order to prove their credibility. To further support their claim, they also link to the bitcoin explorer mentioned above, pointing to transactions that support their case. This increases the distinct incoming link count enormously (The effects of this will be discussed in Section~\ref{sh:networkStructure}), therefore the prioritization scheme favors pages from the scamming subnetwork. This leads to a state where our crawler is only downloading content from a subnetwork of the darknet for hours and not making any progress on the rest of the darknet. Second, about 1.2 million paths have only incoming links from one single hidden service. At this stage a secondary prioritization scheme is required, since the \emph{prioritized} scheme becomes inefficient when it returns 1.2 million candidates to be scheduled for download.
  \item \textbf{Iterative}: This strategy is the result of an attempt to make equal progress on all hidden services as well as on the whole network.  First we make a pass through all hidden services and remember the ones that still have uncrawled paths available. In a second phase we then chose exactly one path to be downloaded per hidden service. In order to select a path for download in the second phase, we use the distinct incoming count as secondary measurement. In case of a draw on the distinct incoming count, we just pick the path we found earlier. This strategy is computationally and I/O intensive, as it needs to do two passes over the data, once over all hidden services and once over all paths. New hidden services are considered as soon as the pool needs to be repopulated. This strategy is then iteratively applied to fill the pool.

\end{itemize}

In the development process, we first started off with the \emph{na\"ive} approach and evolved from there in the order described above. For data collection, we applied a combination of the strategies described. In the beginning of the crawling process, we used the  \emph{new-hidden-service-first} strategy to get an initial broad scan across the available hidden services. We then switched to the \emph{iterative} strategy, since it progressed best on the network. However, we noticed that in the end of the crawling process most hidden services do not have any available paths, and thus at this point the \emph{iterative} strategy is unnecessarily complex. Since the number of paths available per hidden service as well as the distinct incoming counts are in the same order of magnitude for all hidden services,  \emph{random} or  \emph{prioritized} strategy seem more suitable for this phase.

\ifdgruyter
  \begin{figure}[H]
      \centering
      \includegraphics[width=\linewidth]{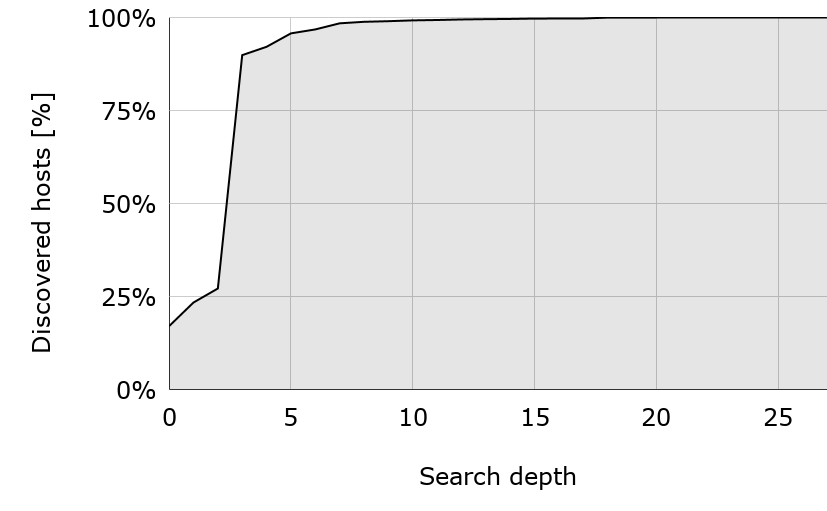}
      \caption{Exploration of hidden services by depth.}
      \label{fig:percentageByDepth}
  \end{figure}
\fi
Running our crawler using these combinations of prioritization schemes, we observe that the connected component of the darknet which we have explored converges sufficiently after 7 hops from our bootstrapping list, as seen in Figure~\ref{fig:percentageByDepth}.

\iflncs
  \begin{wrapfigure}[12]{r}{0.65\textwidth}
      \centering
      \vspace{-55pt}
      \includegraphics[trim={0 1.5cm 0 0}, width=0.65\textwidth]{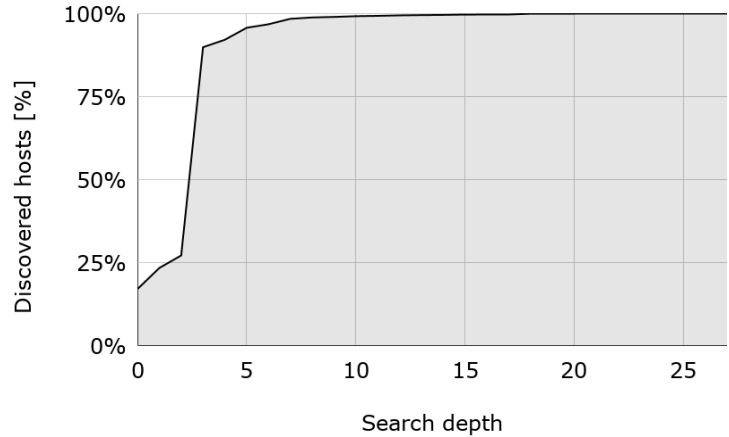}
      \caption{Exploration of hidden services by depth.}
      \label{fig:percentageByDepth}
  \end{wrapfigure}
\fi
\newpage
\subsection{Extracting URIs}

After the content of a website has been downloaded, we need to extract URIs of the downloaded page to understand where it links to and to download further content. Our first approach used regular expressions to match .onion URIs in a string. However, we noticed that URIs can get surprisingly complex, and our regular expression filter was ineffective. We discovered that there exist pathological examples which take several seconds to extract the URIs using such a regular expression. Since we need to extract at least some portion of the URIs fast enough to introduce new URIs in at least the speed data is downloaded, we needed a faster method.

We measured how long it takes to extract links (that are correctly tagged in an HTML document) with cheerio \cite{cheeriojs}, an HTML parsing library. To ensure we only collect links of hidden services, we have a simpler regular expressions ensuring that we do not follow clearnet links. This approach is ten to a hundred times faster than the regular expression-based implementation depending on the input string. However, it has a shortcoming; it only captures links that are part of an \texttt{<a>} tag in the HTML string. Since some pages return some other textual representation or do not correctly link to other pages, we added another component, running the slower regular expression-based technique.

This hybrid approach enables us to keep the crawler running at all times by feeding it with new URIs from the cheerio-based extractor and still have the complete extraction of the regular expression-based extractor.
%
%
\subsection{Storing Darknet Content}
In order to store all the downloaded content as well as the link structure, we use a relational Postgres database. First of all, the \texttt{baseUrls} table stores all the hidden service addresses and a denormalized\footnote{Denormalization is an attempt to increase access speeds by storing the result of a calculation persistently} count of the number of hits, which describes how often this hidden service was referenced.

\iflncs
  \begin{wraptable}[11]{R}[1pt]{9.4cm}
    \vspace{-23pt}
    \begin{tabular}{l|r}
    \hline
    \textbf{Type of Data}           & \textbf{Count} \\
    \hline
    \hline
    Hidden service addresses found      & 34,714                                    \\
    \hline
    Hidden services responding                & 10,957                      \\
    \hline
    Hidden services returning useable content & 7,566                      \\
    \hline
    Paths                           & 7,355,773                   \\
    \hline
    Links                           & 67,296,302                \\
    \hline
    Downloaded paths                & 1,561,590                 \\
    \hline
    Content (useable)               & 667,356         \\
    \hline
    \end{tabular}
    \caption{Detailed numbers of our crawler}
    \label{table:scrapMetrics}
  \end{wraptable}
\fi

We now need the path to specify a full URI: We introduce the \texttt{paths} table, which is keeping the state of an entry. In this role, the paths table contains a flag indicating if a download for this specific path is in progress, a timestamp indicating when the last successful download finished, an integer indicating at which depth in the graph the path was found, the path itself, and a possible subdomain. Additionally, it contains the denormalized count of distinct incoming links to allow for efficient use by our prioritization scheme.

\ifdgruyter
  \begin{table}[H]
      \begin{center}
          \caption{Detailed numbers of our crawler}
          \label{table:scrapMetrics}
          \begin{tabular}[l]{l|r}
          \textbf{Type of Data}           & \textbf{Count} \\
          \hline
          \hline
          Hidden service addresses found      & 34,714                                    \\
          \hline
          Hidden services responding                & 10,957                      \\
          \hline
          Hidden services returning useable content & 7,566                      \\
          \hline
          Paths                           & 7,355,773                   \\
          \hline
          Links                           & 67,296,302                \\
          \hline
          Downloaded paths                & 1,561,590                 \\
          \hline
          Content (useable)               & 667,356         \\
          \end{tabular}
      \end{center}
  \end{table}
\fi

To store the downloaded content, we introduce the \texttt{contents} table, which stores the content itself as well as some meta information about the content such as the HTTP status code and the MIME type. Furthermore, we use a crawl timestamp to track the changes of the network over time while allowing for multiple downloads of the same path.

Last, we introduce the \texttt{link} table that stores the topology of the network. The link table establishes an $n$-to-$m$ mapping between the paths table, since one page may contain multiple links to other pages which can be uniquely identified by their path and the base URL.
%
%
\section{Darknet Structure}
\label{sec:data}
In this section, we present the results of the crawling process and analyze the darknet structure. We show that at least $15\%$ and at most $50\%$ of the darknet is visible. In addition we show that this visible part is well connected. 

\sh{Statistics}
\iflncs
  \begin{wrapfigure}[16]{r}{0.51\textwidth}
    \begin{center}
      \includegraphics[trim={0 0 0 7cm},width=\linewidth]{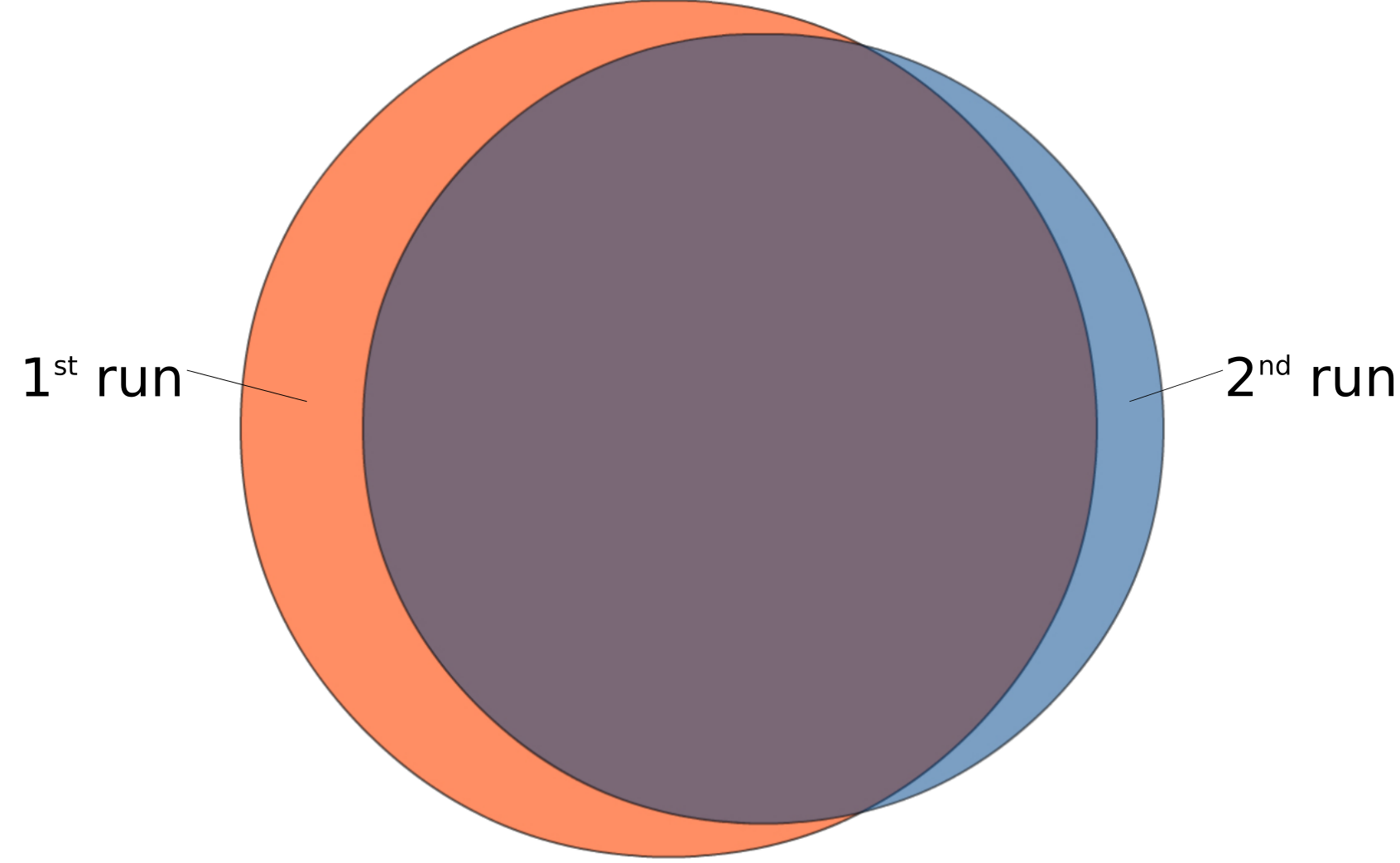}
    \end{center}
    \caption{The Venn diagram shows the difference between two runs a week apart. The first run yielded 10,957 responding hidden services, the second one 9,127. The union of the two contains 11,743 hidden services, the intersection contains 8,341 hidden services}
    \label{fig:highVolatility}
  \end{wrapfigure}
\fi
During the crawling process, we discovered 34,714 hidden service addresses, of which approximately 10,000 hidden services responded to the HTTP(S) request. Out of those hidden services, about 7,500 responded with meaningful content, i.e. not containing an HTTP error status code or an empty HTML page.
We collected in total 7,355,773 paths and 67,296,302 links. Of those, we downloaded 1,561,590 paths. We did not download the other almost 5 million paths because we classified them as black hole paths, highly unlikely containing further URIs to unknown hidden services. The detailed numbers can be found in Table~\ref{table:scrapMetrics}.

The crawler collected data over a time period of one week. According to Tor metrics~\cite{TorMetricsOnion}, the mean of the estimated number of hidden services (darknet hidden services, both visible and invisible) over this time period was 70,640. Having seen 34,714 different hidden services, one could naively assume about half of the darknet hidden services are visible. However, this is only an upper bound, as merely about a third of these responded over the HTTP(S) ports 80 and 443: Either a non-responding hidden service was really not active anymore (and we have just followed a stale hyperlink), or the hidden service decided to ignore our crawler. Having 10,975 responding visible hidden services, we know that at least $15\%$ of the hidden services are visible.
There are multiple reasons why a portion of these hidden services did not respond. One conjecture is that they might connect through other ports or different protocols, or they might discriminate connection requests in some other form.
Such behaviour can be observed in some common link lists, such as the hidden wiki\footnote{Hidden wiki: \texttt{zqktlwi4i34kbat3.onion}}, where a plethora of services offer IRC chat, FTP sharing or P2P sharing and require both a different communication protocol and different ports to be accessed.
\iflncs
  \begin{wraptable}[15]{l}[1pt]{9.4cm}
    \vspace{-10pt}
    \begin{tabular}{l|r}
      \hline
       \textbf{Top Ten MIME Types Returned} & \textbf{Share of Pages} \\
       \hline
       \hline
      text/html     & 85.33\% \\
      \hline
      image/jpeg      & 5.28\% \\
      \hline
      image/png     & 3.23\% \\
      \hline
      text/plain      & 0.85\% \\
      \hline
      application/octet-stream    & 0.72\% \\
      \hline
      application/zip   & 0.55\% \\
      \hline
      application/rss+xml & 0.48\% \\
      \hline
      application/json  & 0.47\% \\
      \hline
      image/gif     & 0.44\% \\
      \hline
      application/epub+zip      & 0.27\% \\
      \hline
    \end{tabular}
    \caption{Top ten MIME types. Text and images contribute about $95\%$ of all content.}
    \label{table:mime_types}
  \end{wraptable}
\fi
\ifdgruyter
  \begin{figure}[H]
      \centering
      \includegraphics[width=0.7\linewidth]{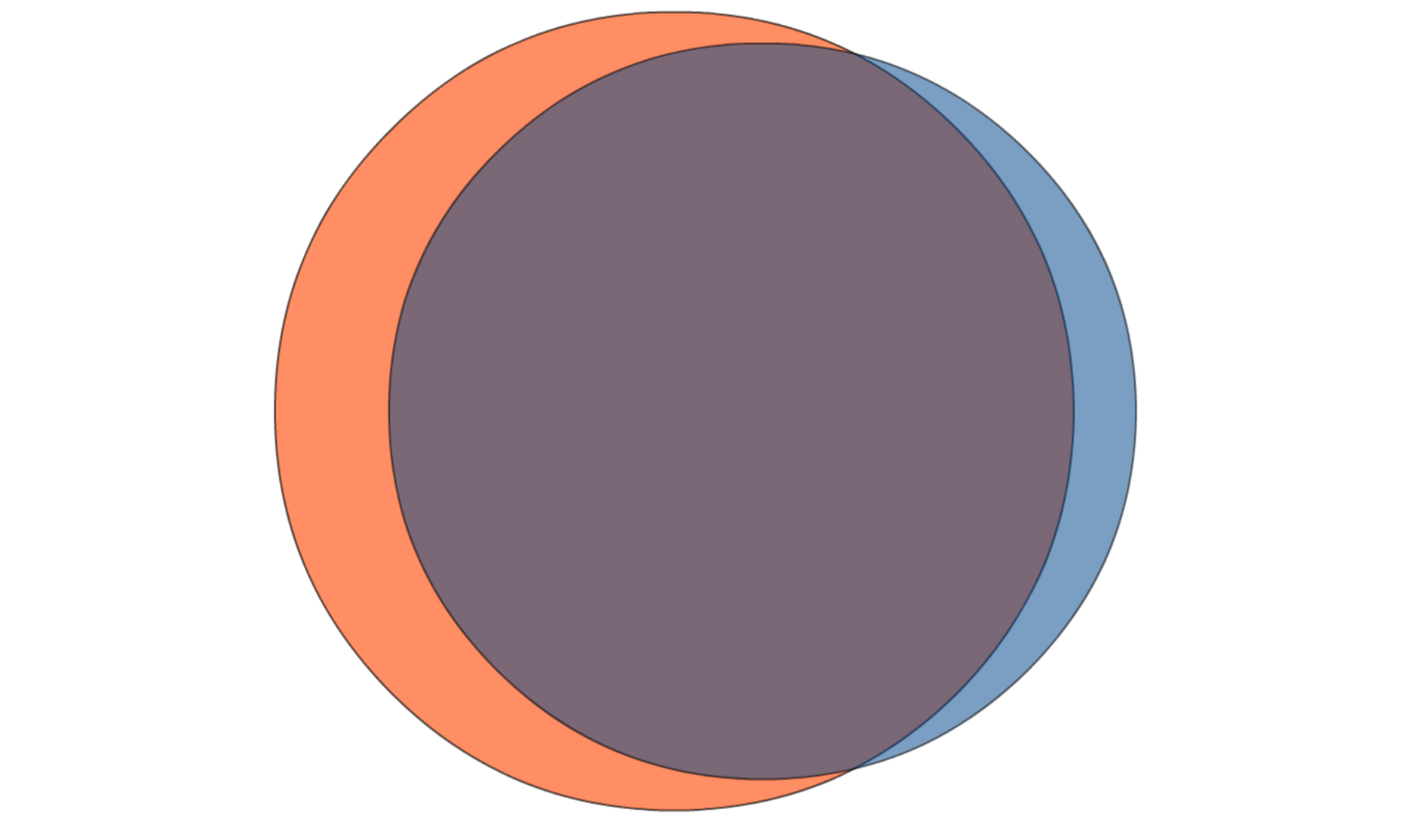}
      \caption{The Venn diagram shows the difference between two runs a week apart. The first run (left circle) yielded 10,957 responding hidden services, the second one (right circle) 9,127. The union of the two contains 11,743 hidden services, the intersection contains 8,341 hidden services}
      \label{fig:highVolatility}
  \end{figure}
\fi

Additionally, we noticed that the network is highly volatile; about $30\%$ of all hidden services running an HTTP(S) server switch from online to offline and vice versa within a week, as depicted in Figure \ref{fig:highVolatility}.
Thus, some of the stale links we followed during the crawling process possibly point to a server that is going to be online again in the future. 

Downloading content of darknet websites that include images or videos is potentially illegal. As we already mentioned, during downloading we detected the MIME type to filter out the potentially illegal content and stored the distribution across MIME types. In total, we found 134 different MIME types, from which the vast majority is of type text/html, as expected. The second most common MIME type we encountered was image types jpeg and png, which is not surprising since almost every webpage has at least an icon that is linked somewhere on the page. The top ten MIME types are listed in the Table \ref{table:mime_types} and a more extensive list of 39 MIME types is available in Appendix \ref{appendix:A}.
\ifdgruyter
  \begin{table}
    \centering
    \caption{Top ten MIME types. Text and images contribute about $95\%$ of all content.}
    \label{table:mime_types}
    \begin{tabular}{l|r}
         \textbf{Top Ten MIME Types Returned} & \textbf{Share of Pages} \\
         \hline
         \hline
    	text/html			&	85.33\% \\
    	\hline
    	image/jpeg			&	5.28\% \\
    	\hline
    	image/png			&	3.23\% \\
    	\hline
    	text/plain			&	0.85\% \\
    	\hline
    	application/octet-stream		&	0.72\% \\
    	\hline
    	application/zip		&	0.55\% \\
    	\hline
    	application/rss+xml	&	0.48\% \\
    	\hline
    	application/json	&	0.47\% \\
    	\hline
    	image/gif			&	0.44\% \\
    	\hline
    	application/epub+zip			&	0.27\% \\
    \end{tabular}
  \end{table}
\fi

%
%
\label{sh:networkStructure}
\sh{Network Structure}
We also studied the network structure (captured links and hidden services) and observed that the darknet behaves like a scale-free network, similarly to the clearnet and various social networks.
In Figure \ref{fig:niceGraphImage}, we present a snapshot of the visible darknet.

\ifdgruyter
  \begin{figure}[H]
      \centering
      \includegraphics[width=0.8\linewidth]{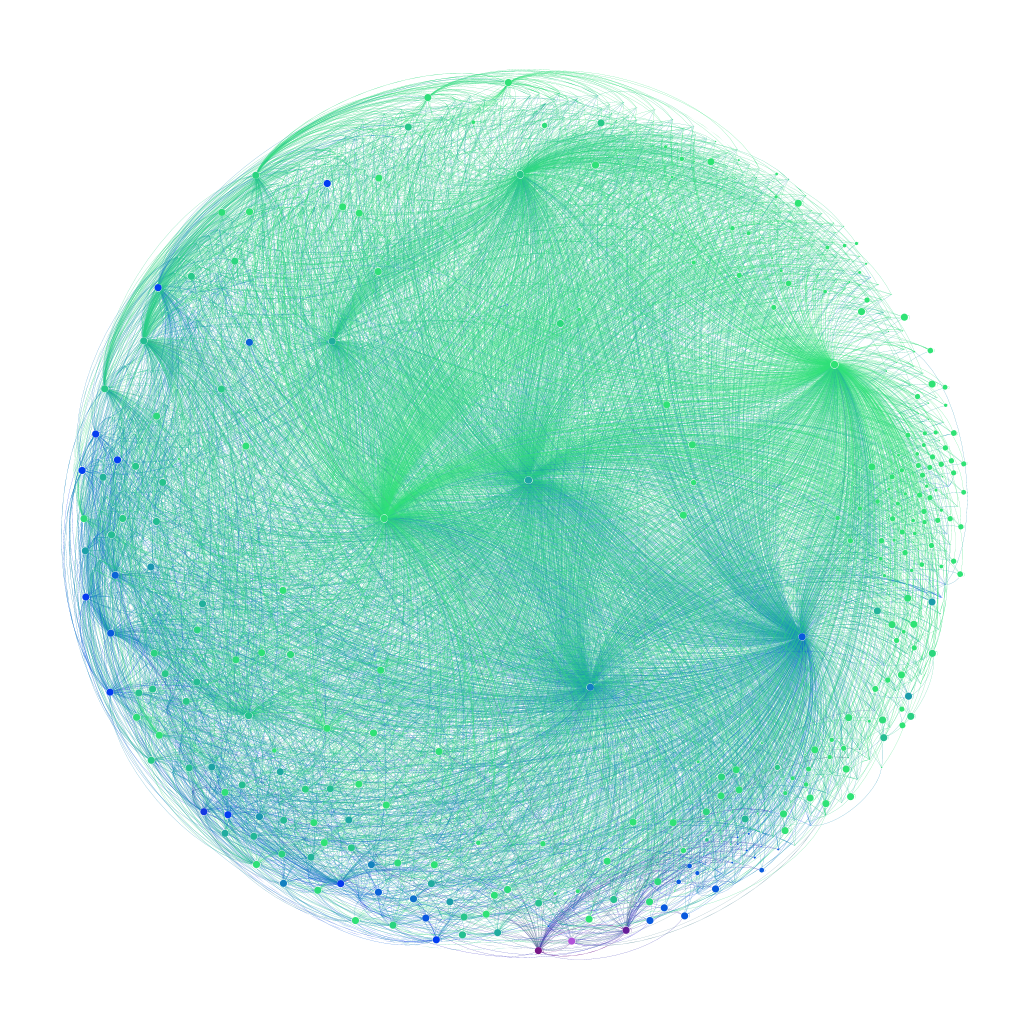}
      \caption{A visualization of approximately $10\%$ of the visible darknet. The size of the node represents how large is the part of the network he has access to. The color indicates the ranking when applying page rank; darker (violet) nodes are the ones with the highest rank.}
      \label{fig:niceGraphImage}
  \end{figure}
\fi
\iflncs
  \begin{figure}[!t]
      \vspace{-50pt}
      \centering
      \includegraphics[width=0.8\linewidth]{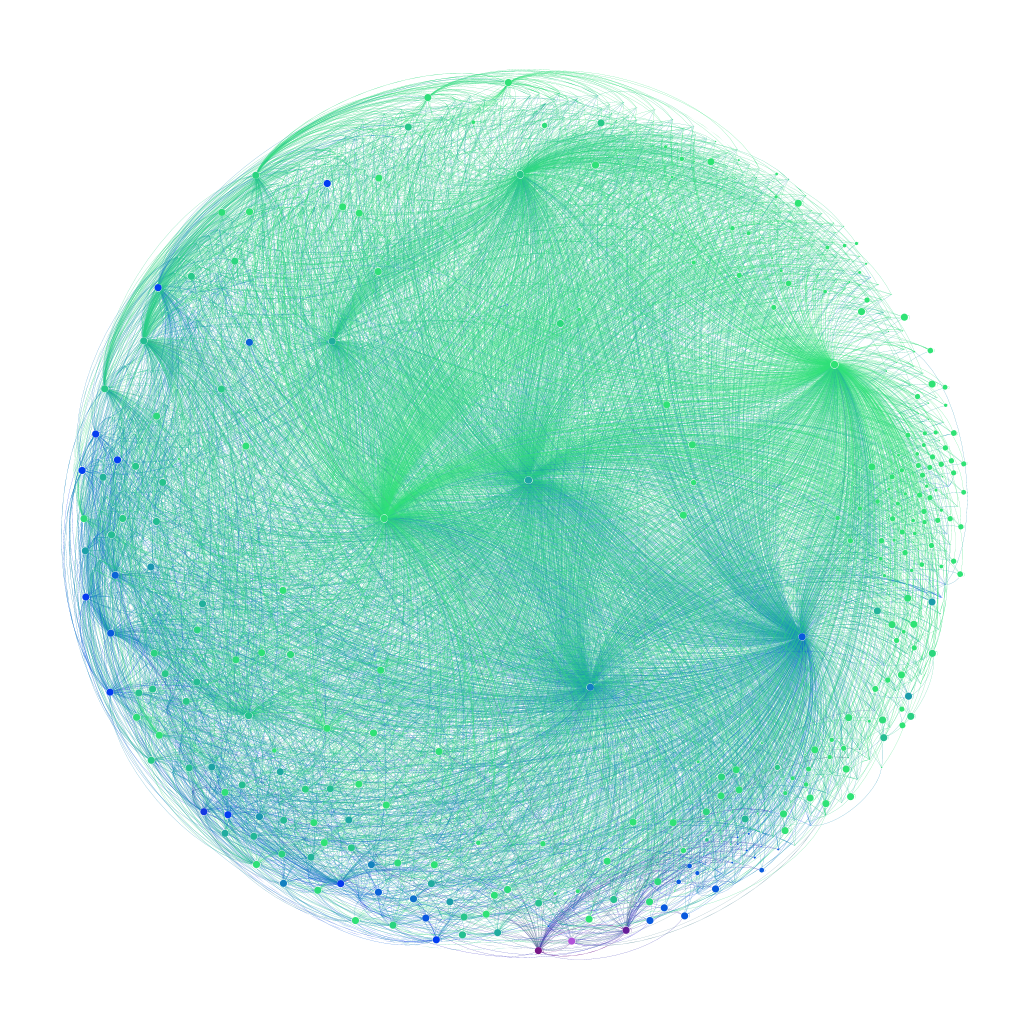}
      \caption{A visualization of approximately $10\%$ of the visible darknet. The size of the node represents how large is the part of the network he has access to. The color indicates the ranking when applying page rank; darker (violet) nodes are the ones with the highest rank.}
      \label{fig:niceGraphImage}
  \end{figure}
\fi

\iflncs
  \begin{wrapfigure}[14]{l}{0.55\textwidth}
  \vspace{-15pt}
  \centering
  \includegraphics[trim={6cm 1.5cm 5cm 1cm}, width=0.65\linewidth]{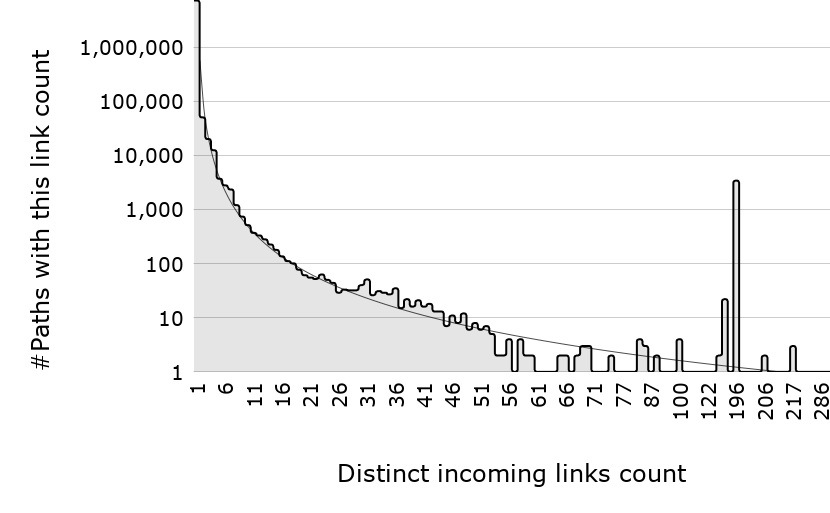}
  \caption{A histogram depicting the count of pages having a particular number of incoming links.}
  \label{fig:powerLawAbnormality}
  \end{wrapfigure}
\fi
The degree distribution of a scale-free network follows a power law and the average path length of the network is proportional to $\ln{N}/\ln{\ln{N}}$. The diameter of the  network is 11. In Figure~\ref{fig:powerLawAbnormality}, it is evident that the degree of the hidden services is power law distributed, except from an outlier which stems from a bitcoing scamming scheme, closely interlinking a high number of hidden services in an attempt to prove their credibility. Moreover, we note that the calculated average path length is $4.1 \approx \frac{\ln 10957}{\ln\ln 10957} = 4.17$. 
Thus, the structure of the visible part of the darknet seems similar to the structure of the clearnet and typical approaches such as crawlers work well on both networks.

\ifdgruyter
  \begin{figure}[H]
  \includegraphics[width=\linewidth]{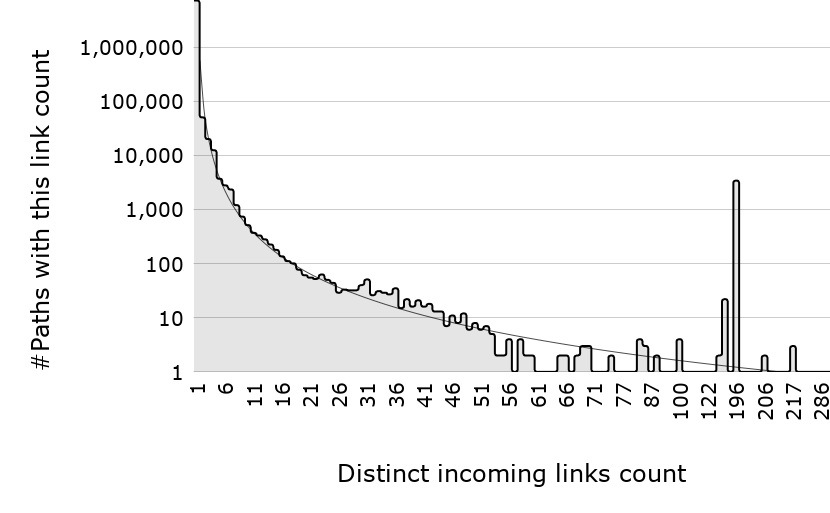}
  \caption{A histogram depicting the count of pages having a particular
               number of incoming links.}
  \label{fig:powerLawAbnormality}
  \end{figure}
\fi

Figure~\ref{fig:powerLawAbnormality} depicts the distribution of incoming link counts. Most paths can only be reached with a single hyperlink. In general the distribution of hyperlinks per path follows a power law distribution. There is an obvious outlier: Several thousand paths have exactly 196 distinct incoming links. This is a bitcoin scamming scheme that tries to increase its credibility by spinning up multiple hidden services, all of which link to each other, that claim that the provided service is valid. Many of these also point to the same bitcoin transactions on a bitcoin explorer site to prove their claim, explaining the second, smaller outlier.

\iflncs
  \begin{wrapfigure}[13]{l}{0.55\textwidth}
      \vspace{-15pt}
      \centering
      \includegraphics[trim={6cm 1.5cm 5cm 1cm}, width=0.62\linewidth]{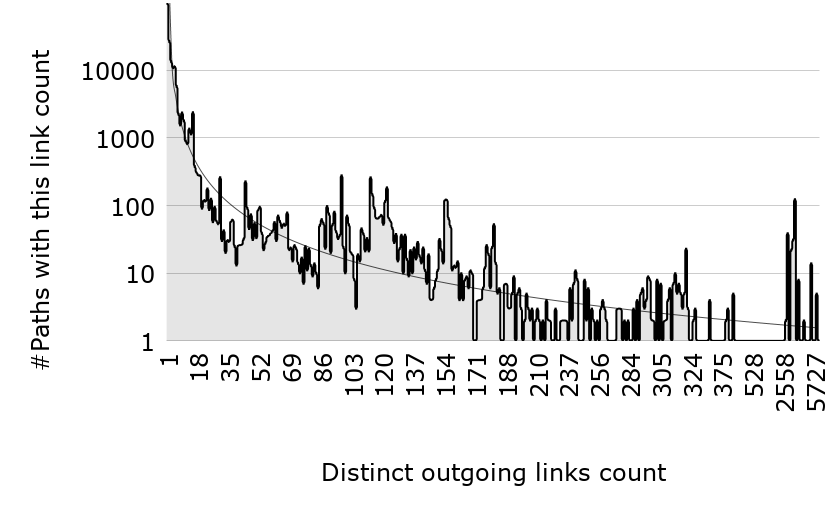}
      \caption{A histogram depicting the count of pages having a particular number of outgoing links.}
      \label{fig:outgoingPowerLaw}
  \end{wrapfigure}
\fi
Comparing Figure~\ref{fig:powerLawAbnormality} with Figure~\ref{fig:outgoingPowerLaw}, we find that measuring the in-degree is much more stable than measuring the out-degree. This is expected to some degree since the incoming links are typically more resistant to manipulation. On the contrary, the outgoing links are fully controlled by the provider of a hidden service, and thus a single page can have arbitrarily many outgoing links.

\ifdgruyter
  \begin{figure}[H]
      \centering
      \includegraphics[width=\linewidth]{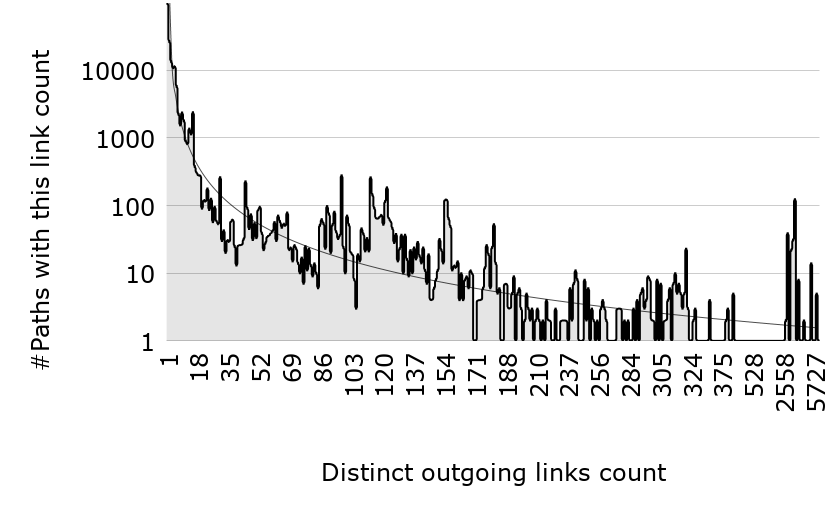}
      \caption{A histogram depicting the count of pages having a particular
               number of outgoing links.}
      \label{fig:outgoingPowerLaw}
  \end{figure}
\fi

\iflncs
  \begin{wrapfigure}[16]{l}{0.57\textwidth}
   \vspace{-15pt}
    \centering
    \includegraphics[trim={4.5cm 1cm 3cm 1cm}, width=0.65\linewidth]{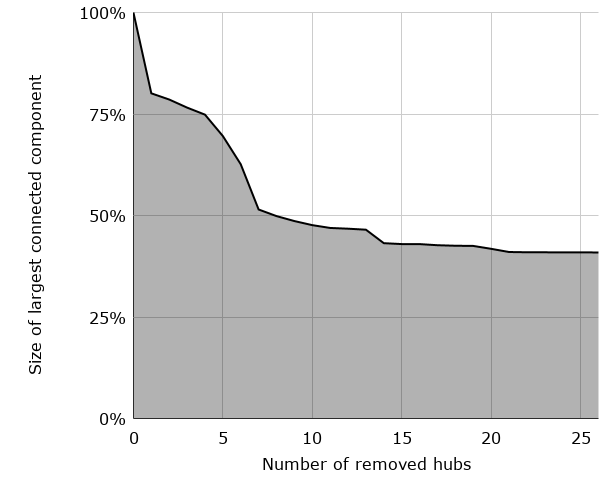}
    \caption{Size of the largest connected component of hidden services after iteratively removing the largest hubs.}
    \label{fig:largestConnectedComponent}
  \end{wrapfigure}
\fi
Our most surprising structural finding is that the visible part of the darknet is well connected. 
While examining the degree distribution, we noticed that the network contains \emph{hubs}, web sites that feature a large number of hyperlinks.  These hubs clearly tie the network together, as they connect sites that do not necessarily have anything in common.
One may expect that removing important hubs will partition the darknet into many small components. However, rather surprisingly, even removing all the largest hubs (with degree greater than 1000) does not disconnect the network.

Figure~\ref{fig:largestConnectedComponent} illustrates the size of the largest connected component after iteratively removing the largest hubs.
\iflncs
  \begin{wraptable}[16]{l}[1pt]{8.5cm}
    \vspace{-20pt}
    \begin{tabular}{l | l | r}
      \hline
      \textbf{Hub Name} & \textbf{URI} & \textbf{Size} \\
      \hline
      \hline
      Fresh onions & zlal32teyptf4tvi.onion  &  7650\\
      \hline
      Onion Land Search & 3bbaaaccczcbdddz.onion  &  7228\\
      \hline
      VisiTOR Search & visitorfi5kl7q7i.onion  &  6210\\
      \hline
      Ahmia & msydqstlz2kzerdg.onion  &  5808\\
      \hline
      H - indexer & jncyepk6zbnosf4p.onion  &  5695\\
      \hline
      List of onion sites & zdxgqrvvvpwnuj2n.onion  &  4045\\
      \hline
      Tor Pastebin & happenedpghnqffa.onion  &  3745\\
      \hline
      Tor Directory & directoryvi6plzm.onion  &  1596\\
      \hline
      The onion crate & cratedvnn5z57xhl.onion  &  1089\\
      \hline
      UnderDir & underdj5ziov3ic7.onion  &  1009\\
      \hline
    \end{tabular}
    \caption{Top ten hub hidden services ranked by the number of distinct \emph{outgoing} links to other services.}\label{table:toptenout}
  \end{wraptable}
\fi
Furthermore, we also examined the hidden services that become disconnected from the largest connected component after removing the largest hubs. It turns out that $90\%$ of those could only be reached through hubs. The remaining $10\%$ form small connected components. In other words, while about half the visible darknet is well-connected even without hubs, the other half are single sites or small clusters of sites that are only reachable through hubs. 
\ifdgruyter
  \begin{table}[t]
      \begin{center}
          \caption{Top ten hub hidden services ranked by the number of distinct \emph{outgoing} links to other services.}
          \label{table:toptenout}
          \begin{tabular}{p{3cm} | p{3.5cm} | p{1cm}}
               \textbf{Hub Name} & \textbf{URI} & \textbf{Size} \\
               \hline
               \hline
              Fresh onions & zlal32teyptf4tvi.onion  &  7650\\
              \hline
              Onion Land Search & 3bbaaaccczcbdddz.onion  &  7228\\
              \hline
              VisiTOR Search & visitorfi5kl7q7i.onion  &  6210\\
              \hline
              Ahmia & msydqstlz2kzerdg.onion  &  5808\\
              \hline
              H - indexer & jncyepk6zbnosf4p.onion  &  5695\\
              \hline
              List of onion sites & zdxgqrvvvpwnuj2n.onion  &  4045\\
              \hline
              Tor Pastebin & happenedpghnqffa.onion  &  3745\\
              \hline
              Tor Directory & directoryvi6plzm.onion  &  1596\\
              \hline
              The onion crate & cratedvnn5z57xhl.onion  &  1089\\
              \hline
              UnderDir & underdj5ziov3ic7.onion  &  1009\\
          \end{tabular}
      \end{center}
  \end{table}
\fi

\ifdgruyter
  \begin{table}[b]
      \centering
      \begin{tabular}{l|l|r}
          \textbf{Hub Name}   & \textbf{URI}  & \textbf{Size} \\
          \hline
          \hline
        DreamMarket           & [Multiple]          & 2130 \\
        \hline
        Daniel's Hosting        & dhosting4okcs22v.onion    & 889 \\
        \hline
        Blockchain Explorer       & blockchainbdgpzk.onion    & 523 \\
        \hline
        The Tor Mail Box        & torbox3uiot6wchz.onion    & 280 \\
        \hline
        \hbox{\strut DuckDuckGo}
        (Search engine)
        & 3g2upl4pq6kufc4m.onion    & 166 \\
        \hline
        Deep Dot Web (News)       & deepdot35wvmeyd5.onion    & 151 \\
        \hline
        \hbox{\strut Matrix Trilogy}
        (Image sharing)
        & matrixtxri745dfw.onion    & 139 \\
        \hline
        TORCH: Tor Search       & xmh57jrzrnw6insl.onion    & 134 \\
        \hline
        Hidden Wiki           & zqktlwi4fecvo6ri.onion    & 126 \\
        \hline
        Imperial Library (Books)    & xfmro77i3lixucja.onion    & 124 \\
      \end{tabular}
      \caption{Top ten hidden services ranked by the number of \emph{incoming} links from unique other hidden services.}
      \label{table:toptenin}
  \end{table}
\fi
Some large hubs also feature a search engine. Among the top ten hubs, presented in Table \ref{table:toptenout}, there are three search engines, six lists of links which also download the title of the website to provide some sort of searchable content, and one Pastebin site where links are deposited without organization. Note that the linked hidden services can contain disturbing content (text or images), as some include previews of the linked pages. Additionally some of the URLs provided in this paper might not work anymore due to the high volatility of the Tor network. 

In Table~\ref{table:toptenin}, we present the top ten hidden services ranked by the number of incoming links from unique hidden services. 
First, we observe that one market place is clearly dominant in this list. 
Moreover, we notice that the second most linked hidden service provides hosting and related services for darknet pages for free. In exchange many of the pages running on this hoster's infrastructure link back to the hidden service. 
The third most linked hidden service is the same blockchain explorer that breaks the \emph{random} strategy, as mentioned in subsection \ref{sec:prioritizationDownload}. 
We found that the bitcoin scamming scheme that breaks the \emph{prioritized} strategy links to this particular bitcoin explorer. The bitcoin scamming scheme makes up for more than $50\%$ of the distinct incoming links of the bitcoin explorer.
However, the most remarkable observation is that a library made it into the list of the most referenced services.

\ifdgruyter
  \begin{figure}[H]
  \includegraphics[width=\linewidth]{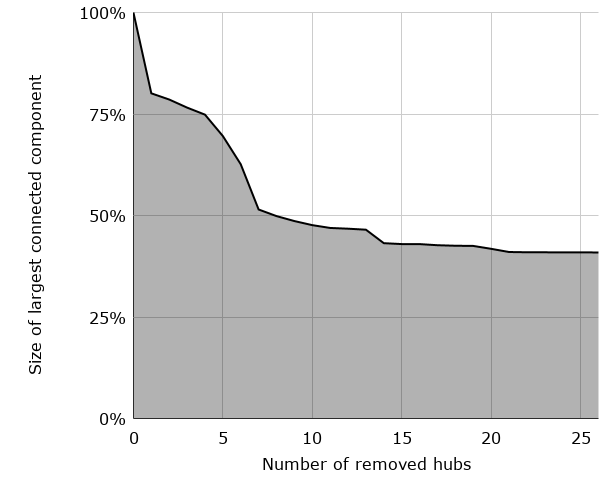}
  \caption{Size of the largest connected component of hidden services after iteratively removing the largest hubs.}
  \label{fig:largestConnectedComponent}
  \end{figure}
\fi

\iflncs
  \begin{table}[H]
    \begin{tabular*}{\textwidth}{l@{\extracolsep{\fill}}l r}
      \hline
        \textbf{Hub Name}   & \textbf{URI}  & \textbf{Size} \\
        \hline
        \hline
      DreamMarket           & [Multiple]          & 2130 \\
      \hline
      Daniel's Hosting        & dhosting4okcs22v.onion    & 889 \\
      \hline
      Blockchain Explorer       & blockchainbdgpzk.onion    & 523 \\
      \hline
      The Tor Mail Box        & torbox3uiot6wchz.onion    & 280 \\
      \hline
      \hbox{\strut DuckDuckGo}
      (Search engine)
      & 3g2upl4pq6kufc4m.onion    & 166 \\
      \hline
      Deep Dot Web (News)       & deepdot35wvmeyd5.onion    & 151 \\
      \hline
      \hbox{\strut Matrix Trilogy}
      (Image sharing)
      & matrixtxri745dfw.onion    & 139 \\
      \hline
      TORCH: Tor Search       & xmh57jrzrnw6insl.onion    & 134 \\
      \hline
      Hidden Wiki           & zqktlwi4fecvo6ri.onion    & 126 \\
      \hline
      Imperial Library (Books)    & xfmro77i3lixucja.onion    & 124 \\
      \hline
    \end{tabular*}
    \vspace{10pt}
    \caption{Top ten hidden services ranked by the number of \emph{incoming} links from unique other hidden services.}
    \label{table:toptenin}
  \end{table}
\fi
\section{Analysis Methodology}
Given the size of the collected data set of 667,356 paths, we relied on automated preprocessing and machine learning techniques to classify their contents. The main modules used to achieve this are:
\begin{itemize}
    \item A \emph{data preprocessor} to normalize the downloaded content.
    \item A \emph{classifier} that classifies the contents based on the preprocessed intermediate representations.
    \item A \emph{database} to hold both the intermediate representation as well as the classified results.
\end{itemize}

The goal of the classifier is twofold: Firstly, to determine the legality of a particular piece of content, which we express as a binary decision. Secondly, to classify the content into a particular category describing its topic. The modules we developed to perform these tasks are part of our open source release and are described in detail below.
%
%
\subsection{Content Preprocessing}
The preprocessor has two main tasks to accomplish: First it linguistically preprocesses the contents and then builds the indexes needed for classification. 
Initially, we extract the clean text from the content.
For this purpose, we use cheerio~\cite{cheeriojs} to strip any HTML tags and other script tags we find in the input string.
After getting a clean string, we identify the physical language of the content, in order to decide which linguistic preprocessing will be applied to the string.
We employ franc~\cite{WormerFranc} and the minimal set of 82 languages franc provides. Franc also offers more inclusive language sets; however, a more inclusive language set might produce worse language classification results, since it is too fine grained (e.g., Scottish English vs. British English).

We found that approximately $66\%$ of the content is in English, followed by $11\%$ in Russian. The physical language distribution is depicted in Figure~\ref{fig:languages}. In total we found 60 different language. For a more extensive list of 26 languages, see Appendix \ref{appendix:B}. Since the vast majority of content is written in English, we solely consider English language in the classification process.

\ifdgruyter
  \begin{figure}[H]
  \includegraphics[width=\linewidth]{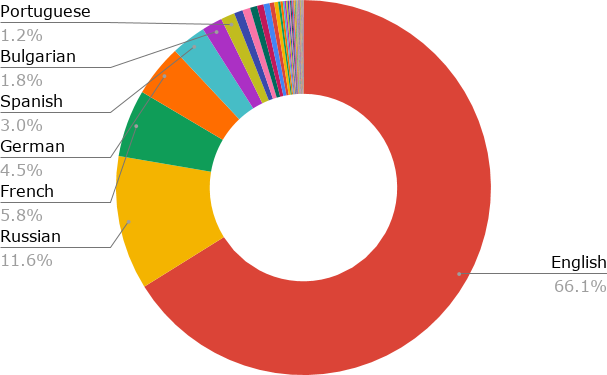}
  \caption{Language distribution in the visible darknet.}
  \label{fig:languages}
  \end{figure}
\fi

\iflncs
  \begin{figure}[H]
  \centering
  \includegraphics[width=0.6\linewidth]{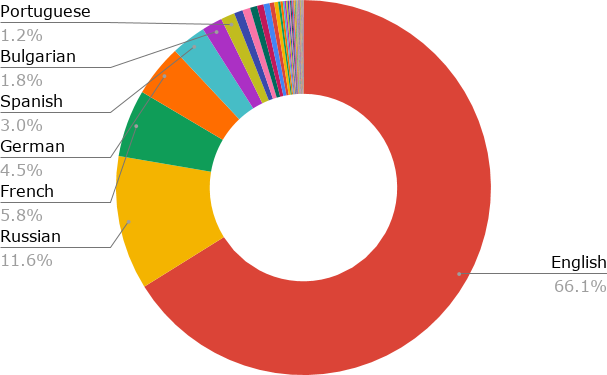}
  \caption{Language distribution in the visible darknet.}
  \label{fig:languages}
  \end{figure}
\fi

For the following steps we need an array of terms instead of a string; thus we tokenize the content string into words.
Now that the language of the content is known, we remove stop words~\cite{McDowall} depending on the language. As it turns out, some of the content only contains stop words, resulting in an empty content after stop-words removal. This content is immediately marked as \emph{empty}.

After stop word removal, we normalize the content further and apply stemming \cite{WormerStemmer} based on the porter stemmer \cite{porter1980algorithm} when appropriate. Previous work \cite{Harman1991,hull1996stemming,krovetz1996word,SaltonTextProcessing} suggests that stemming does not work equally well in all languages, hence we only apply stemming for English content. Lemmatization (determining the canonical form of a set of words depending on the context) is not considered since previous work showed that it can even degrade performance when applied to English content.

Next, we build a positional index, which allows us to choose whether the classification should happen on a list of words, a bag of words or a set of words approach. Since we want to keep the most information possible and employ an SVM classifier, we decided to use a bag of words approach for the classification of labels. However, the software provides the functionality for all three approaches, which makes it easier to experiment with different machine learning techniques on the collected data set. For the bag of words model, we need to store the frequency of each word; the frequency can be extracted from the positional index, which is discussed in the following subsection. 
In addition, we keep the extracted clean content string to facilitate manual classification. Thereby, the clean content string can be directly presented to a human classifier, in order to get the labeled training data set.
%
%
\subsection{Storing Content for Classification}
We store the extracted content in the table \texttt{cleanContents} with a foreign key to the original content. This table has a one-to-one assignment for both the \texttt{language} as well as the primary class label to the \texttt{labels} table. Since the legality of a document is expressed as a binary decision, we directly use a boolean column to store the legal value, where \emph{true} corresponds to legal, and \emph{false} to illegal.

The goal of the training phase is to employ active learning~\cite{Xu2009}; hence, it is essential to measure the certainty of the classifier's decisions. For this purpose, we insert a column to store the probability output of the classifier for both label and legal classification. Now, we can apply active learning techniques in the classification phase.

A positional index stores information about which term appears in which document at which positions. The positional index itself is constructed with three different tables. The first table is the list of \texttt{terms}, which has a one-to-many mapping to the \texttt{postings} table. A posting is a pair, mapping a term to a document. Therefore each posting in the posting table has a foreign relation to a clean content and a term, thus creating the tuple $<$\texttt{term} ID, \texttt{cleanContent} ID$>$. In order to store a positional index, every posting points to multiple \texttt{positions}, describing the positions at which the referenced term appears in the specified clean content.

%
%
\subsection{Content Classification}
As discussed in \cite{Nabki2017,Verma2013}, a Support Vector Machine (SVM) is well suited for text classification, mainly because the SVM efficiently copes with high dimensional features that are typical for text classification. Moreover, an SVM is easy to train and apply, compared to different types of neural networks. We use $k$-fold cross-validation and compare the output scores to decide which parameters and kernels are performing better. The parameters that deliver the best results are then used to train the model.

We begin the classification process with a very small training set of about $200$ entries, in an attempt to minimize the size of the training set, and then apply active learning~\cite{Xu2009}. 
Active learning requires knowledge on the uncertainty level of the classifier for each sample, since in every iteration we manually label the samples with the highest uncertainty. The idea is to  manually label the entries with the highest entropy to convey the most information to the classifier.
 
 These entries are then added to the data set, and the previously trained classifier is trained again on the updated data set. Since the original training set was small, we reevaluated different SVM classifiers from time to time to ensure we are still using the best possible configuration. We found that a linear kernel worked best for both classification problems. With this approach, we reached an F1 score of $60\%$ and an accuracy of $75\%$ for the label classifier, and an F1 score of $85\%$ and an accuracy of $89\%$ for the legal classifier.

During  the training of the classifier and the examination of preliminary results, we observed that some pages did not contain enough information to be assigned a label. In this case, the \emph{empty} label was added and excluded from the corpus. Pages not containing enough information were either actually (or almost) empty, often containing only a placeholder like ``coming soon'', or were login pages without further information about what a user is logging into or for what service the user could register. If there are other pages of the same hidden service available, it is sufficient to assign the page a label. If no other pages are available, or they do not contain enough information for labeling, we are not able to classify such a hidden service.

The labels are chosen in an attempt to represent the most common categories and answer often posed questions about content found in the Tor network.
We always labeled the contents with the most specific label possible. For example, if we have a page that is a blog about information security, this page would be labeled ``Security''.

In order to label the hidden services, we take the maximum number of labeled pages per hidden service and assign this label to the hidden service as well. For legal classification, we assign \emph{illegal} to a hidden service as soon as at least $10\%$ of the contents are illegal. This threshold is used to reduce false positives from the classification process. In Figure~\ref{fig:legalityByThreshold}, we show how the number of illegal hidden services changes with different thresholds. Lower values than $1\%$ only introduce minor changes to the result. Overall, we observe that  the number of illegal hidden services is quite stable.

\ifdgruyter
  \begin{figure}[H]
    \centering
    \includegraphics[width=0.8 \linewidth]{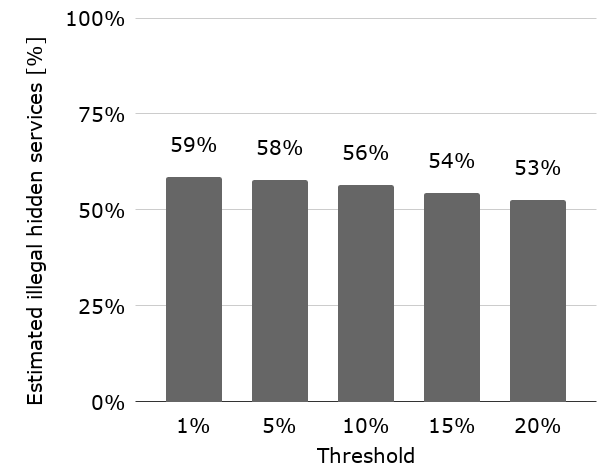}
    \caption{Estimated number of hidden services containing illegal content for different thresholds.}
    \label{fig:legalityByThreshold}
  \end{figure}
\fi

\iflncs
  \begin{figure}[H]
    \centering
    \includegraphics[width=0.55\linewidth]{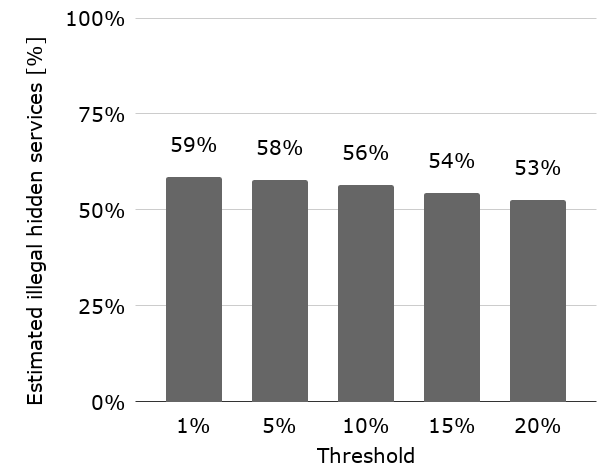}
    \caption{Estimated number of hidden services containing illegal content for different thresholds.}
    \label{fig:legalityByThreshold}
  \end{figure}
\fi

The labels are listed below, each with a description and examples.
\begin{itemize}
	\item \textbf{Activism}: Any content that is conveying activism, e.g. human rights activism.
	\item \textbf{Adult}: Pornographic content of any type except paedophilic content. It does not have to contain media content to be classified as Adult; textual descriptions  are also included if they are clearly of pornographic nature.
	\item \textbf{Art}: Anything ranging from literature to images (either photographs or computer generated) to whole web pages being an art project. This label does not include pages that contain sharing books or other pieces of art. Such a page would belong to the Sharing label.
	\item \textbf{Blog}: Any web page that has a clearly defined set of authors that can publish content on the page and the content is not generated interactively as in a forum. Any web page, even if not in typical blog form is counted here if it fulfills the above condition and cannot be labeled more specifically, 
	including marketing or news content.
	\item \textbf{Counterfeit}: Any page that offers or describes how to create counterfeits is listed here. Most common topics are passports and credit cards.
	\item \textbf{Drugs}: Any page where the main topic is drugs. This does not only include marketplaces selling drugs but also forums and information pages about drugs, including drug prevention pages.
	\item \textbf{Empty}: The page does not contain enough information to be labeled with some other label. This category mainly includes empty pages, pages with too short messages and login pages that have no reference to the service they belong to. Furthermore, pages containing an error code (e.g. \texttt{404}) in the returned HTML were also assigned the Empty label.
	\item \textbf{Exchange}: Any currency exchange page, typically between cryptocurrencies or from cryptocurrency to fiat money and back.
	\item \textbf{Forum}: Either forums or chats. We consider them to belong to the same label since in both cases users are populating the page with content and responding to each other. The only difference is of a temporal nature, how fast one expects a response. This label is only assigned when no other label fits better. For example, a forum focused on drugs would be classified into the Drugs label.
	\item \textbf{Fraud}: Offers that are of a fraudulent nature, such as bitcoin frauds that attempt to trick the user into sending them bitcoin for some promised higher return value.
	\item \textbf{Gambling}: Mainly online casinos running with cryptocurrencies or other betting games that promise to pay out money for certain events.
	\item \textbf{Hacking}: Contains anything related to breaching computer systems, independent of being black hat or white hat hacking.
	\item \textbf{Hosting}: Offers to host a web page, darknet as well as clearnet.
	\item \textbf{Laundry}: Pages offering money laundering services, often for cryptocurrencies stemming from dishonest sources.
	\item \textbf{Hub}: Hubs are pages listing links of darknet sites, such as ``The Hidden Wiki''.
	\item \textbf{Mail}: E-mail provider, anonymized or not.
	\item \textbf{Market}: Any market selling goods that do not qualify for other labels. Also markets that sell multiple goods of different types.
	\item \textbf{Murder}: Pages promising to execute a murder in exchange for money. Some of these pages could also be considered scams as indicated in news and blog publications \cite{Monteiro}. However, we introduce this label separately since hitmen services are often linked to the darknet.
	\item \textbf{Paedophilic}: Any pages containing either paedophilic content directly, including media content or textual representation of fantasies. This label also contains forums and support groups for paedophiles.
	\item \textbf{Phishing}: Fake pages that try to grab login credentials for different services from the user (e.g. Facebook or Google).
	\item \textbf{Search Engine}: Any page offering to search for darknet hidden services. Often these keep a manually edited index.
	\item \textbf{Security}: Pages containing information about computer security and personal security.
	\item \textbf{Seized}: Pages taken offline by law enforcement agencies.
	\item \textbf{Service}: Any service that does not fit into one of the other labels, e.g. a darknet calendar.
	\item \textbf{Sharing}: Any page that shares content, either illegal (pirating) or legal, e.g., Dropbox.
	\item \textbf{Social Network}: Social networks similar to Facebook or Twitter. Facebook itself is accessible through the darknet~\cite{facebook} and thus belongs to this label.
	\item \textbf{Software Development}: Anything that has to do with software or its development. We consolidated selling software and development into one label since in most cases, the page serves as both the marketing page as well as the git repository or test page.
	\item \textbf{Terrorism}: Anything related to extremism and terrorism. We do not classify the type of terrorism.
	\item \textbf{Weapons}: Selling or discussing goods around weapons, be it guns, rifles or other devices that can kill or harm people.
	\item \textbf{Whistleblowing}: Pages similar to Wikileaks. Pages that contain a drop off for whistle-blowers also qualify for this label.
\end{itemize}

The darknet seems to be a shelter for people discussing controversial or embarrassing topics. One such example is a support group for paedophiles that tries to help its members to live with their predisposition without harming others and avoid interference with the law. This example shows that it is not enough to just assign one of the previous mentioned labels to assess the contents legality, since in most categories both legal and illegal content can be found. Therefore, we also trained a binary classifier to determine the legality of content. The classifier also uses an SVM for classification and the same bag-of-words data collection as the label classifier.

It is important to note that assessing the legality of content is challenging for several reasons. For one, every country and legal system has different notions of which actions are legal and illegal.
Furthermore, sometimes the content itself belongs in a grey area, between legal and illegal. Since our goal is to provide a rough estimate on the distribution of the content, we follow a rather conservative approach when assessing legality: when the purpose of the content is an obviously illegal action, the content is classified as illegal, even if it might not technically be. One might say that this classifier is slightly biased towards classifying contents as illegal.
\section{Darknet Content}
\label{sec:results}

In this section, we present the results of the content classification, both in terms of label and legality. 
The visible darknet mostly consists of empty pages; just above $30\%$ of the content is classified in this label. 
For the legal classification of both pages and hidden services, we exclude the content with an empty label. Such content usually does not contain enough information for labeling, as we cannot assess the intended content from a vanilla login page; also regarding legality, empty pages are impossible to judge, and hence not considered.
\ifdgruyter
  \begin{figure}[h]
  \includegraphics[width=\linewidth]{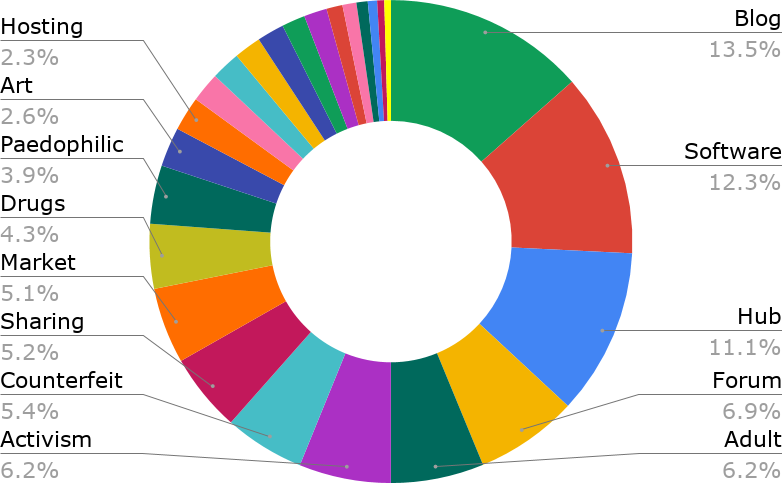}
  \caption{Label of darknet pages}
  \label{fig:labelsByContent}
  \end{figure}
\fi

\iflncs
  \begin{figure}[h]
  \centering
  \includegraphics[width=0.6\linewidth]{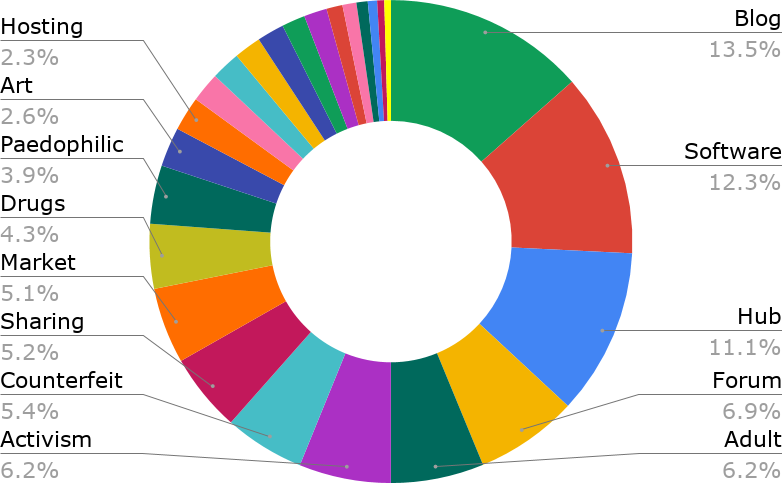}
  \caption{Label of darknet pages}
  \label{fig:labelsByContent}
  \end{figure}
\fi

\ifdgruyter
  \begin{figure}[h]
  \includegraphics[width=\linewidth]{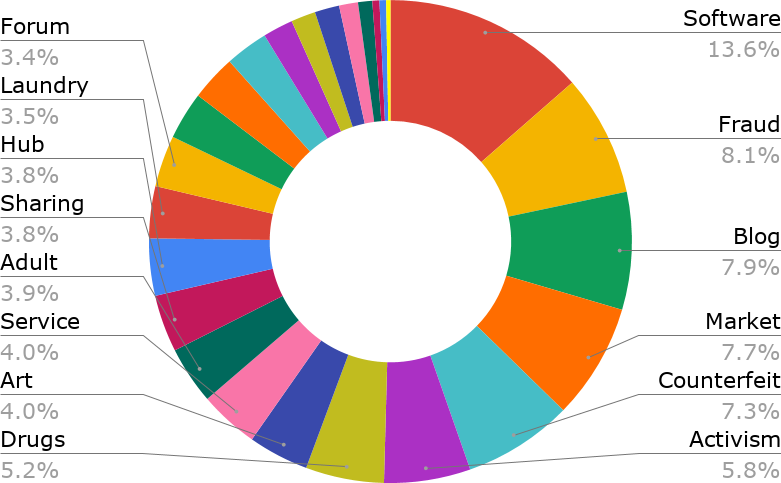}
  \caption{Label of darknet hidden services}
  \label{fig:labelsByHost}
  \end{figure}
\fi

\iflncs
  \begin{figure}[h]
  \centering
  \includegraphics[width=0.6\linewidth]{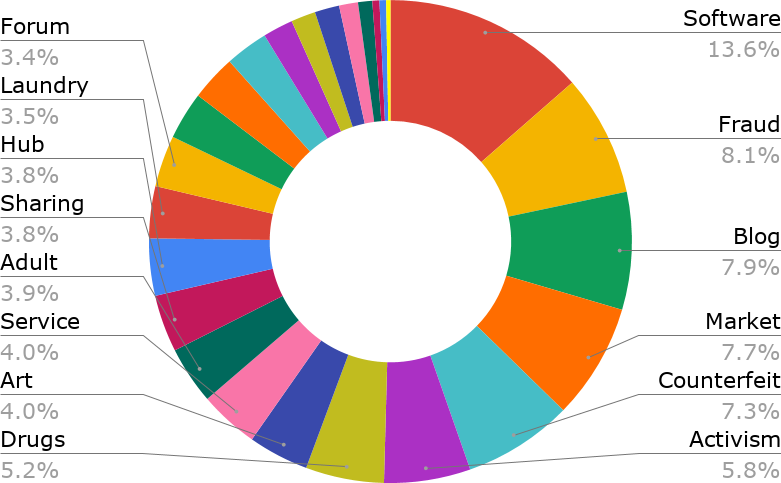}
  \caption{Label of darknet hidden services}
  \label{fig:labelsByHost}
  \end{figure}
\fi
\newpage
\sh{Label Classification}
In Figure \ref{fig:labelsByContent}, we present the top thirteen labels of darknet pages, while the complete list can be found in Appendix~\ref{appendix:labelsByPages}. 
We notice that, quite unexpectedly, \emph{Blog} is the top category in the classification per page. However, as we discussed earlier, we classified in \emph{Blog} all the pages that did not fit in any other label, such as a company blog presenting the new employees or a news article about an arbitrary topic.

On the other hand, the classification per hidden service was more challenging since typically most hidden services have multiple pages with sometimes different labels. To deal with this issue we decided on the following classification strategy: we counted how many pages each hidden service hosts from each label and assigned to the hidden service the label with the highest count. Consolidating all the labeled data per hidden service, as depicted in Figure \ref{fig:labelsByHost}, we observe only minor changes in the relative occurrence of labels. The complete list can be found in Appendix~\ref{appendix:labelsByHosts}.

The top category in classification per hidden service is \emph{Software}, which also seems surprising. 
Exploring further, however, we observed a specific pattern which justifies the high occurrence of \emph{Software} labels:
first of all, many pages rely on open source software, which is later developed further. To manage such developments, the pages introduce a version control system which is often hosted by the same hidden service as the actual service. In addition, many pages store information for the operation team in status information pages which are also classified in the \emph{Software} category. Both version control systems and operation tools often contain a large amount of sub pages, so \emph{Software} may be overrepresented. 
Furthermore, the way we accumulate the labels of single pages in order to label the hidden services also contributes to this result. A different feature selection could probably improve the results and allow the classifier to distinguish between the main topic of a hidden service and the git repository of such.

If we compare the results by page and by hidden service, we see two major differences. Fraud is only making up for $1.82\%$ when grouped by page, however, it is the second largest category when grouping by hidden services with $8.13\%$. We find that many hidden services that try to scam their users only contain very few sub pages and rely mainly on external, more trusted services to place their links and faked testimonies.
In contrast, hubs make up for $11.1\%$ of all pages, while they only constitute $3.8\%$ of all hidden services. We observe that many link lists have a high amount of sub pages, as to be expected.

\sh{Legal Classification}
After excluding all empty pages, we find that about $60\%$ of the collected pages is legal, as illustrated in Figure \ref{fig:legalByContent}; this seems an unexpectedly high number for the darknet. To explain this high number of legal pages, we look more closely into the assigned labels. 
For example, \emph{Software} is often considered legal because the development of software such as a market place is legal; however the goods sold in such a marketplace might be illegal. Thus, the label we assigned might be misleading in assessing the purpose of the page.

\ifdgruyter
  \begin{figure}[H]
    \includegraphics[width=\linewidth]{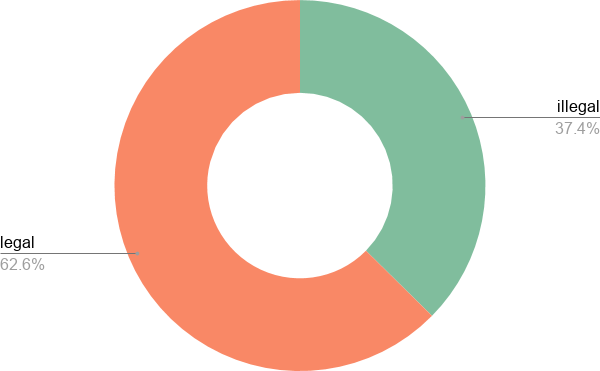}
    \caption{Approximately $60\%$ of all pages are legal in the visible part of the darknet.}
    \label{fig:legalByContent}
  \end{figure}
\fi

\iflncs
  \begin{figure}[H]
    \centering
    \includegraphics[width=0.6\linewidth]{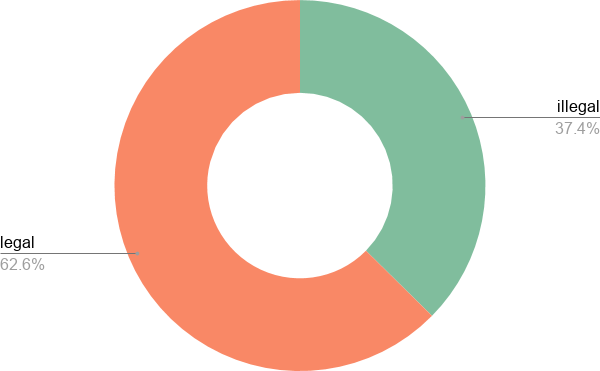}
    \caption{Approximately $60\%$ of all pages are legal in the visible part of the darknet.}
    \label{fig:legalByContent}
  \end{figure}
\fi

\ifdgruyter
  \begin{figure}[H]
    \includegraphics[width=\linewidth]{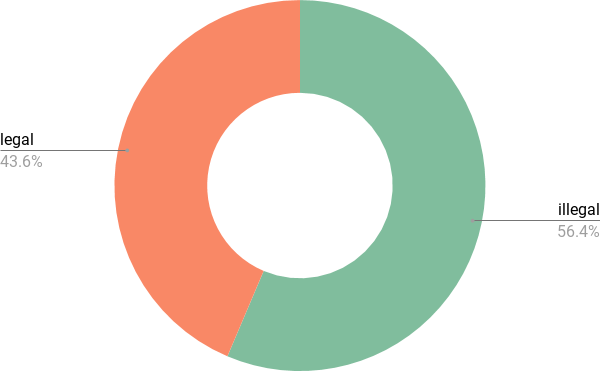}
    \caption{When accumulating the pages by hidden service, we find that above $50\%$ of all hidden services contain illegal content.}
    \label{fig:legalByHost}
  \end{figure}
\fi

\iflncs
  \begin{figure}[H]
    \centering
    \includegraphics[width=0.6\linewidth]{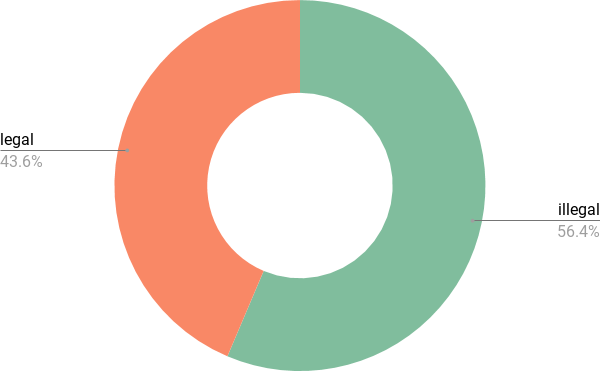}
    \caption{When accumulating the pages by hidden service, we find that above $50\%$ of all hidden services contain illegal content.}
    \label{fig:legalByHost}
  \end{figure}
\fi

If we accumulate the labels by hidden service, the number of hidden services containing illegal content slightly increases compared to the corresponding number of pages, as depicted in Figure \ref{fig:legalByHost}.
However, we found that a lot of forums have dedicated threads for the illegal content to enable users to choose in which discussion to participate. When classifying the content of the forum, some part contributes to the legal and some to the illegal category. However, we classify the hidden service as illegal if at least $10\%$ of the content is illegal. Thus, the classification by hidden service is biased towards the illegal category; this is the most probable cause of the percentage difference between pages and hidden services regarding legality.


Note that all the results apply to content written in English only, c.f., Figure~\ref{fig:languages}.

\section{Related Work}\label{sec:relatedwork}

Starting with the rise of the web, crawling has been a topic of interest  for the academic community as well as society in general. Some of the earliest crawlers \cite{Gray1993,McBryan1994,Eichmann1994,Pinkerton1994} were presented just shortly after the web became public. However, since the early web was smaller than the visible darknet today, clearnet crawlers did not face the same issues as they do today. Castillo~\cite{Castillo2005} and Najork and Heydon~\cite{Najork2002} addressed a wide variety of issues, e.g. scheduling. Castillo found that applying different scheduling strategies works best. Both acknowledged the need to limit the number of requests sent to the targets, and described several approaches similar to ours.

Contrary to aforementioned work, we are interested in exploring the \emph{darknet}.
Previous work~\cite{GeorgeKadianakis2015,TorMetricsOnion} has estimated the size of the Tor darknet using non-invasive methods that allow us to measure what portion of the darknet we have explored.
Similarly to our work, Al Nabki et al.~\cite{Nabki2017} want to understand the Tor network; they created a dataset of active Tor hidden services and explored different methods for content classification. However, instead of using a real crawler, they browsed a big initial dataset as seed, and then followed all links they found exactly twice. In turn, Moore and Rid~\cite{Moore2016} went to depth five, which seems to be good enough to catch most of the darknet, as their results are similar to ours. Our crawler does a real recursive search and as such understands the darknet topology in more detail, in contrast to both of these previous works.

Other work applied different approaches to data collection for classification.
Owen et al.~\cite{Owen2016} measured the activity on each page of the darknet by operating a large number of Tor relays, a measure which is not possible with our methodology. Similarly to our work, Biryukov et al.~\cite{Biryukov2014} have classified darknet content with findings comparable to ours. Their methodology relied on a Tor bug which was fixed in 0.2.4.10-alpha and can no longer be used for data collection. Unlike both of these works which can have adverse effects on the anonymity of Tor users, our method is less invasive as it only utilizes publicly available information and does not require operating any Tor relays or participating in the DHT.

Furthermore, Terbium labs~\cite{ClareGollnick2016} proposed content classification via a small but random sample from a crawled collection. Despite the small size of the dataset, their results are also comparable to ours.

While we focus on the web, Chaabane et al.~\cite{Chaabane2010} wanted to know what other protocols are using Tor, by analyzing the traffic on Tor's exit routers. They concluded that the largest amount of transmitted data stems from peer-to-peer file sharing traffic. As shown by McCoy et al.~\cite{McCoy2008}, measurements of traffic can be easily skewed by data-heavy applications such as file sharing. This work estimates that HTTP makes up for approximately $92\%$ of all connections. 

Content classification is also a topic in the clearnet, and depending on the size of the collected dataset, automated content classification is required. 
Both Verma~\cite{Verma2013} and Kim and Kuljis~\cite{Kim2010} analyzed existing techniques and proposed different methods to improve content analysis to Web based content. Xu et al.~\cite{Xu2009} compared different methods to improve SVM performance, including the active learning approach. They showed that an active learning approach coupled with an SVM classifier can result in better accuracy with less samples than other compared methods. These results were applied in the machine learning process we employed, combining active learning and SVM estimators.

Specifically for the darknet, Fidalgo et al. \cite{Fidalgo2017} described how to classify illegal activity on the darknet based on image classification. While adding images as an additional feature could improve the performance of our classifier (especially for pages that did not contain enough textual information), we abstained from it because  it requires downloading potentially illegal images.


A more specific content analysis was performed on marketplaces residing on the darknet. Christin et al. \cite{Christin2013} and Soska et al. \cite{KyleSoska2015} concentrated on a few large markets and developed strategies to circumvent barriers such as captchas and password entries. They used this data to estimate sales volumes, product categories and activity of vendors.

\section{Conclusion $\&$ Future Work}

We performed a thorough analysis of both the structure and the content of the visible part of the darknet. To the best of our knowledge, we were the first to systematically (recursively) explore the network topology of the visible darknet. To that end, we developed a novel open source spider which captures the graph of hyperlinks between web pages. 

Remarkably, we found that the darknet is well connected and behaves like a scale-free network, similarly to the clearnet. The degree distribution follows a power law and the diameter of the network is 11. Furthermore, we identified the existence of large hubs connecting the network. Although these hubs have  high degrees, their impact on the connectivity of the network is as significant as anticipated; at least half of the darknet remains well connected even if we remove all such hubs. Last but not least, we observed that the darknet is highly volatile, and experience a churn of about $30\%$ within a single week.

We employed supervised learning techniques to classify the content of the darknet both per page and per hidden service. Specifically, we used active learning and an SVM classifier to label the content in terms of physical language, topic and legality. 
Surprisingly, 60\% of the pages and approximately half of the hidden services in the darknet are legal. Furthermore, the label classification indicates that (contrary to popular belief) about half of the pages in the darknet concern licit topics such as blogs, forums, hubs, activism etc. On the other hand, we revealed that a significant amount of hidden services perform fraud. 

Our crawler was able to uncover a significant but small ($15\%$) percentage of the darknet as compared to Tor Metrics estimates. Future work can address the reason we were unable to penetrate the rest of the darknet. We already recommended several mechanisms towards this direction, i.e. looking for non-HTTP(S) services that could be running on the darknet as well as employing the crawler repeatedly for a longer period of time to increase the probability of discovering services which are often offline.

The machine learning techniques can also be improved to obtain better results. In particular, we classified hidden services into labels by taking a majority of labels of their individual pages. However, treating hidden services independently could probably yield better classification results.  
Furthermore, taking into account additional features in the classification process could significantly improve the outcome. Explicitly handling special cases such as git repositories and excluding them from the accumulation of the labels could have significant impact on the results. Moreover, notable improvement could arise from the use of images as an additional classification feature when the content is insufficient; however, this method can only be adopted if the legality of the image can be automatically determined.

Lastly, our collected data is comprehensive enough to be useful if made available to the public. To that end, future work could build an open source Tor search engine that allows searching the indexed data, similarly to rudimentary closed source solutions such as TORCH~\cite{torch}. Our indexing techniques could also benefit existing open source search engine attempts such as Ahmia~\cite{ahmia}. An interactively browseable version of our graph that presents an in-depth apprehensible view on the topology of the darknet could also be a valuable extension for future work.

\newpage

\iflncs
    \bibliographystyle{abbrv}
    \bibliography{references}
\fi
\ifdgruyter
    \printbibliography
\fi
\appendix
\newpage
%
%
\section{Appendix: MIME Types}
\label{appendix:A}
MIME Types of collected content. Either the server returned a MIME type header or we detected the type by data inspection through mmmagic\cite{mscdex}
\iflncs
\begin{center}
\label{table:mime types}
\begin{tabular*}{\textwidth}{l@{\extracolsep{\fill}}r}
    \textbf{MIME Type} & \textbf{Share of pages}\\
	\hline
	\hline
	text/html			&	85.33\% \\
	\hline
	image/jpeg			&	5.28\% \\
	\hline
	image/png			&	3.23\% \\
	\hline
	text/plain			&	0.85\% \\
	\hline
	application/octet-stream		&	0.72\% \\
	\hline
	application/zip		&	0.55\% \\
	\hline
	application/rss+xml	&	0.48\% \\
	\hline
	application/json	&	0.47\% \\
	\hline
	image/gif			&	0.44\% \\
	\hline
	application/epub+zip			&	0.27\% \\
	\hline
	application/xml		&	0.25\% \\
	\hline
	application/javascript			&	0.25\% \\
	\hline
	application/x-fictionbook+xml	&	0.20\% \\
	\hline
	application/x-bittorrent		&	0.19\% \\
	\hline
	application/atom+xml			&	0.19\% \\
	\hline
	application/x-mpegURL			&	0.18\% \\
	\hline
	text/xml			&	0.17\% \\
	\hline
	application/pdf		&	0.15\% \\
	\hline
	text/css			&	0.15\% \\
	\hline
	audio/x-wav			&	0.13\% \\
	\hline
	application/x-download			&	0.08\% \\
	\hline
	video/mp4			&	0.08\% \\
	\hline
	video/webm			&	0.08\% \\
	\hline
	application/vnd.comicbook+zip	&	0.05\% \\
	\hline
	text/csv			&	0.03\% \\
	\hline
	audio/mpeg			&	0.02\% \\
	\hline
	application/rdf+xml	&	0.02\% \\
	\hline
	application/pgp-signature		&	0.02\% \\
	\hline
	image/jpg			&	0.01\% \\
	\hline
	text/tab-separated-values		&	0.01\% \\
	\hline
	application/x-tar	&	0.01\% \\
	\hline
	application/vnd.comicbook-rar	&	0.01\% \\
	\hline
	audio/x-scpls		&	0.01\% \\
	\hline
	text/calendar		&	0.01\% \\
	\hline
	application/owl+xml	&	0.01\% \\
	\hline
	video/x-matroska	&	0.01\% \\
	\hline
	image/svg+xml		&	0.01\% \\
	\hline
	application/x-empty	&	0.01\% \\
	\hline
	Other				&	0.06\% \\
	\hline
\end{tabular*}
\end{center}
\fi

\ifdgruyter
\begin{center}
\label{table:mime types}
\begin{supertabular}[H]{l|r}
    \textbf{MIME Type} & \textbf{Share of pages}\\
	\hline
	\hline
	text/html			&	85.33\% \\
	\hline
	image/jpeg			&	5.28\% \\
	\hline
	image/png			&	3.23\% \\
	\hline
	text/plain			&	0.85\% \\
	\hline
	application/octet-stream		&	0.72\% \\
	\hline
	application/zip		&	0.55\% \\
	\hline
	application/rss+xml	&	0.48\% \\
	\hline
	application/json	&	0.47\% \\
	\hline
	image/gif			&	0.44\% \\
	\hline
	application/epub+zip			&	0.27\% \\
	\hline
	application/xml		&	0.25\% \\
	\hline
	application/javascript			&	0.25\% \\
	\hline
	application/x-fictionbook+xml	&	0.20\% \\
	\hline
	application/x-bittorrent		&	0.19\% \\
	\hline
	application/atom+xml			&	0.19\% \\
	\hline
	application/x-mpegURL			&	0.18\% \\
	\hline
	text/xml			&	0.17\% \\
	\hline
	application/pdf		&	0.15\% \\
	\hline
	text/css			&	0.15\% \\
	\hline
	audio/x-wav			&	0.13\% \\
	\hline
	application/x-download			&	0.08\% \\
	\hline
	video/mp4			&	0.08\% \\
	\hline
	video/webm			&	0.08\% \\
	\hline
	application/vnd.comicbook+zip	&	0.05\% \\
	\hline
	text/csv			&	0.03\% \\
	\hline
	audio/mpeg			&	0.02\% \\
	\hline
	application/rdf+xml	&	0.02\% \\
	\hline
	application/pgp-signature		&	0.02\% \\
	\hline
	image/jpg			&	0.01\% \\
	\hline
	text/tab-separated-values		&	0.01\% \\
	\hline
	application/x-tar	&	0.01\% \\
	\hline
	application/vnd.comicbook-rar	&	0.01\% \\
	\hline
	audio/x-scpls		&	0.01\% \\
	\hline
	text/calendar		&	0.01\% \\
	\hline
	application/owl+xml	&	0.01\% \\
	\hline
	video/x-matroska	&	0.01\% \\
	\hline
	image/svg+xml		&	0.01\% \\
	\hline
	application/x-empty	&	0.01\% \\
	\hline
	Other				&	0.06\% \\
	\hline
\end{supertabular}
\end{center}
\fi

\newpage
%
%
\section{Appendix: Languages in the Darknet}
\label{appendix:B}
We found content in 60 languages. The exact number of documents per language is listed below.
\ifdgruyter
\begin{center}
    \label{table:languages}
    \begin{supertabular}[H]{l|r}
    \textbf{Language} &	\textbf{Share of pages}	\\
    \hline
    \hline
	English		&	66.14\%\\
	\hline
	Russian		&	11.56\%\\
	\hline
	French		&	5.82\%\\
	\hline
	German		&	4.49\%\\
	\hline
	Spanish		&	3.00\%\\
	\hline
	Bulgarian	&	1.78\%\\
	\hline
	Portuguese	&	1.20\%\\
	\hline
	Italian	&	0.75\%\\
	\hline
	Mandarin	&	0.68\%\\
	\hline
	Iloko		&	0.61\%\\
	\hline
	\textit{Undetermined}	&	0.56\%\\
	\hline
	Dutch		&	0.53\%\\
	\hline
	Swedish		&	0.40\%\\
	\hline
	Ukrainian	&	0.34\%\\
	\hline
	Turkish		&	0.20\%\\
	\hline
	Polish		&	0.20\%\\
	\hline
	Hausa		&	0.14\%\\
	\hline
	Madurese	&	0.14\%\\
	\hline
	Korean		&	0.13\%\\
	\hline
	Standard Arabic	&	0.11\%\\
	\hline
	Japanese	&	0.10\%\\
	\hline
	Czech		&	0.10\%\\
	\hline
	Romanian	&	0.10\%\\
	\hline
	Malay		&	0.08\%\\
	\hline
	Indonesian	&	0.08\%\\
	\hline
	\textit{Other}		&	0.79\%\\
    \end{supertabular}
\end{center}
\fi
\iflncs
\begin{center}
    \label{table:languages}
    \begin{tabular*}{\textwidth}{l@{\extracolsep{\fill}}r}
    \textbf{Language} &	\textbf{Share of pages}	\\
    \hline
    \hline
	English		&	66.14\%\\
	\hline
	Russian		&	11.56\%\\
	\hline
	French		&	5.82\%\\
	\hline
	German		&	4.49\%\\
	\hline
	Spanish		&	3.00\%\\
	\hline
	Bulgarian	&	1.78\%\\
	\hline
	Portuguese	&	1.20\%\\
	\hline
	Italian	&	0.75\%\\
	\hline
	Mandarin	&	0.68\%\\
	\hline
	Iloko		&	0.61\%\\
	\hline
	\textit{Undetermined}	&	0.56\%\\
	\hline
	Dutch		&	0.53\%\\
	\hline
	Swedish		&	0.40\%\\
	\hline
	Ukrainian	&	0.34\%\\
	\hline
	Turkish		&	0.20\%\\
	\hline
	Polish		&	0.20\%\\
	\hline
	Hausa		&	0.14\%\\
	\hline
	Madurese	&	0.14\%\\
	\hline
	Korean		&	0.13\%\\
	\hline
	Standard Arabic	&	0.11\%\\
	\hline
	Japanese	&	0.10\%\\
	\hline
	Czech		&	0.10\%\\
	\hline
	Romanian	&	0.10\%\\
	\hline
	Malay		&	0.08\%\\
	\hline
	Indonesian	&	0.08\%\\
	\hline
	\textit{Other}		&	0.79\%\\
    \end{tabular*}
\end{center}
\fi

\newpage
\section{Appendix: Complete List of Labels}
All labelled categories are listed here in detail.
\subsection{Labels by Pages}
\label{appendix:labelsByPages}
\ifdgruyter
\begin{center}
    \begin{supertabular}[H]{l|r}
    \textbf{Label}		&	\textbf{Share of Pages}\\
	\hline
	\hline
	Blog		&	13.49\%\\
	\hline
	Software Development	&	12.29\%\\
	\hline
	Hub		&	11.14\%\\
	\hline
	Forum		&	6.86\%\\
	\hline
	Adult		&	6.23\%\\
	\hline
	Activism	&	6.16\%\\
	\hline
	Counterfeit	&	5.37\%\\
	\hline
	Sharing		&	5.24\%\\
	\hline
	Market		&	5.08\%\\
	\hline
	Drugs		&	4.32\%\\
	\hline
	Pedophilic	&	3.91\%\\
	\hline
	Art			&	2.64\%\\
	\hline
	Hosting		&	2.30\%\\
	\hline
	Service		&	1.99\%\\
	\hline
	Hacking		&	1.95\%\\
	\hline
	Fraud		&	1.82\%\\
	\hline
	Gambling	&	1.81\%\\
	\hline
	Whistleblowing			&	1.57\%\\
	\hline
	Search Engine			&	1.54\%\\
	\hline
	Laundry		&	1.07\%\\
	\hline
	Exchange	&	0.94\%\\
	\hline
	Seized		&	0.75\%\\
	\hline
	Security	&	0.62\%\\
	\hline
	Murder		&	0.47\%\\
	\hline
	Social Network			&	0.44\%\\
    \end{supertabular}
\end{center}
\fi

\iflncs
\begin{center}
    \begin{supertabular*}{\textwidth}{l@{\extracolsep{\fill}}r}
    \textbf{Label}		&	\textbf{Share of Pages}\\
	\hline
	\hline
	Blog		&	13.49\%\\
	\hline
	Software Development	&	12.29\%\\
	\hline
	Hub		&	11.14\%\\
	\hline
	Forum		&	6.86\%\\
	\hline
	Adult		&	6.23\%\\
	\hline
	Activism	&	6.16\%\\
	\hline
	Counterfeit	&	5.37\%\\
	\hline
	Sharing		&	5.24\%\\
	\hline
	Market		&	5.08\%\\
	\hline
	Drugs		&	4.32\%\\
	\hline
	Pedophilic	&	3.91\%\\
	\hline
	Art			&	2.64\%\\
	\hline
	Hosting		&	2.30\%\\
	\hline
	Service		&	1.99\%\\
	\hline
	Hacking		&	1.95\%\\
	\hline
	Fraud		&	1.82\%\\
	\hline
	Gambling	&	1.81\%\\
	\hline
	Whistleblowing			&	1.57\%\\
	\hline
	Search Engine			&	1.54\%\\
	\hline
	Laundry		&	1.07\%\\
	\hline
	Exchange	&	0.94\%\\
	\hline
	Seized		&	0.75\%\\
	\hline
	Security	&	0.62\%\\
	\hline
	Murder		&	0.47\%\\
	\hline
	Social Network			&	0.44\%\\
    \end{supertabular*}
\end{center}
\fi

\subsection{Labels by Hidden Services}
\label{appendix:labelsByHosts}
\ifdgruyter
\begin{center}
    \begin{supertabular}[H]{l|r}
    \textbf{Label}  & \textbf{Share of Hidden Service}\\
    \hline
    \hline
    Software Development	&	13.56\%\\
	\hline
	Fraud		&	8.13\%\\
	\hline
	Blog		&	7.87\%\\
	\hline
	Market		&	7.75\%\\
	\hline
	Counterfeit	&	7.33\%\\
	\hline
	Activism	&	5.79\%\\
	\hline
	Drugs		&	5.25\%\\
	\hline
	Art			&	4.05\%\\
	\hline
	Service		&	3.98\%\\
	\hline
	Adult		&	3.86\%\\
	\hline
	Sharing		&	3.83\%\\
	\hline
	Hub		&	3.83\%\\
	\hline
	Laundry		&	3.48\%\\
	\hline
	Forum		&	3.45\%\\
	\hline
	Whistleblowing			&	3.19\%\\
	\hline
	Hosting		&	3.07\%\\
	\hline
	Hacking		&	2.88\%\\
	\hline
	Search Engine			&	2.02\%\\
	\hline
	Pedophilic	&	1.64\%\\
	\hline
	Gambling	&	1.64\%\\
	\hline
	Exchange	&	1.26\%\\
	\hline
	Seized		&	0.95\%\\
	\hline
	Murder		&	0.47\%\\
	\hline
	Security	&	0.44\%\\
	\hline
	Social Network			&	0.28\%\\
    \end{supertabular}
\end{center}
\fi

\iflncs
\begin{center}
    \begin{supertabular*}{\textwidth}{l@{\extracolsep{\fill}}r}
    \textbf{Label}  & \textbf{Share of Hidden Service}\\
    \hline
    \hline
    Software Development	&	13.56\%\\
	\hline
	Fraud		&	8.13\%\\
	\hline
	Blog		&	7.87\%\\
	\hline
	Market		&	7.75\%\\
	\hline
	Counterfeit	&	7.33\%\\
	\hline
	Activism	&	5.79\%\\
	\hline
	Drugs		&	5.25\%\\
	\hline
	Art			&	4.05\%\\
	\hline
	Service		&	3.98\%\\
	\hline
	Adult		&	3.86\%\\
	\hline
	Sharing		&	3.83\%\\
	\hline
	Hub		&	3.83\%\\
	\hline
	Laundry		&	3.48\%\\
	\hline
	Forum		&	3.45\%\\
	\hline
	Whistleblowing			&	3.19\%\\
	\hline
	Hosting		&	3.07\%\\
	\hline
	Hacking		&	2.88\%\\
	\hline
	Search Engine			&	2.02\%\\
	\hline
	Pedophilic	&	1.64\%\\
	\hline
	Gambling	&	1.64\%\\
	\hline
	Exchange	&	1.26\%\\
	\hline
	Seized		&	0.95\%\\
	\hline
	Murder		&	0.47\%\\
	\hline
	Security	&	0.44\%\\
	\hline
	Social Network			&	0.28\%\\
    \end{supertabular*}
\end{center}
\fi

\iflncs
\vspace*{\fill}
\noindent This work is licensed under the Creative Commons Attribution 4.0 International License. To view a copy of this license, visit \url{http://creativecommons.org/licenses/by/4.0/} or send a letter to Creative Commons, PO Box 1866, Mountain View, CA 94042, USA.
\fi

\end{document}